\documentstyle[12pt]{article}
\input epsf
\begin{document}
\baselineskip=15pt
\newcommand{\x}{{\bf x}}
\newcommand{\y}{{\bf y}}
\newcommand{\z}{{\bf z}}
\newcommand{\bp}{{\bf p}}
\newcommand{\A}{{\bf A}}
\newcommand{\B}{{\bf B}}
\newcommand{\p}{\varphi}
\newcommand{\del}{\nabla}
\newcommand{\be}{\begin{equation}}
\newcommand{\ee}{\end{equation}}
\newcommand{\bq}{\begin{eqnarray}}
\newcommand{\eq}{\end{eqnarray}}
\newcommand{\ba}{\begin{eqnarray}}
\newcommand{\ea}{\end{eqnarray}}
\def\r{\nonumber\cr}
\def\hf{\textstyle{1\over2}}
\def\qr{\textstyle{1\over4}}
\def\Sc{Schr\"odinger\,}
\def\sc{Schr\"odinger\,}
\def\'{^\prime}
\def\>{\rangle}
\def\<{\langle}
\def\-{\rightarrow}
\def\dbd{\partial\over\partial}
\def\tr{{\rm tr}}
\def\hg{{\hat g}}
\def\ca{{\cal A}}
\def\pd{\partial}
\def\dl{\delta}
\newcommand{\n}{\nonumber\\}
\def\k2{\kappa^2}
\def\km2{\kappa^{-2}}
\def\lm{\lambda}
\def\sg{\sigma}
\def\al{\alpha}
\def\gm{\gamma}
\def\Gm{\Gamma}
\def\om{\omega}
\def\tt{\theta}

\def\hf{{1 \over 2}}
\def\qt{{1 \over 4}}
\def\tr{{\rm Tr}}
\def\tP{{\tilde P}}
\def\th{{\tilde h}}
\def\hg{{\hat g}}
\def\hph{{\hat \phi}}
\def\tph{{\tilde \phi}}
\def\hR{{\hat R}}
\def\hA{{\hat A}}
\def\hB{{\hat B}}
\def\bxi{{\bar \xi}}
\def\cD{{\cal D}}
\def\hh{{\hat h}}
\def\hV{{\hat V}}
\def\hzt{{\hat \zeta}}

\begin{titlepage}
\vskip1in
\begin{center}
{\large The Boundary Weyl Anomaly in the ${\cal N}=4$ SYM/Type IIB
Supergravity Correspondence.}
\end{center}
\vskip1in
\begin{center}
{\large Paul Mansfield$^a$, David Nolland$^b$ and Tatsuya Ueno$^b$

\vskip20pt

$^a$Department of Mathematical Sciences

University of Durham

South Road

Durham, DH1 3LE, England

{\it P.R.W.Mansfield@durham.ac.uk}

\vskip20pt

$^b$Department of Mathematical Sciences

University of Liverpool

Liverpool, L69 3BX, England

{\it nolland@liv.ac.uk}

{\it ueno@liv.ac.uk} }

\end{center}
\vskip1in
\begin{abstract}

\noindent We give a complete account of the Schr\"odinger
representation approach to the calculation of the Weyl anomaly of
${\cal N}=4$ SYM from the AdS/CFT correspondence. On the AdS side,
the $1/N^2$ correction to the leading order result receives
contributions from all the fields of Type IIB Supergravity, the
contribution of each field being given by a universal formula. The
correct matching with the CFT result is thus a highly non-trivial
test of the correspondence.

\end{abstract}

\end{titlepage}

\section{Introduction: the basic calculation}

When Super-Yang-Mills theory is coupled to a non-dynamical,
external metric, $g_{ij}$, the Weyl anomaly, ${\cal A}$, is the
response of the `free-energy' (i.e. the logarithm of the partition
function), $F$, to a scale transformation of that metric. This is
a quantum effect, since the classical theory is scale-invariant,
but the one-loop result is exact because supersymmetry prevents
higher loops from contributing. The result is proportional to
$N^{2}-1¥$ and Henningson and Skenderis showed \cite{Henningson}
that the $N^{2}¥$ part is correctly reproduced by a tree-level
calculation in five-dimensional gravity confirming the Maldacena
conjecture to leading order in large $N$. Reproducing quantum
effects in a gauge theory from classical gravity is itself truly
remarkable but to go beyond the leading order and reproduce the
$-1$ requires much more than just classical gravity, it needs the
computation of Superstring loops. This is a stringent test of the
details of the conjecture because although the graviton alone is
responsible for the leading order result, all the species of
fields in IIB Supergravity contribute at subleading order. The
purpose of this paper is to show that the subleading contribution
to the Weyl anomaly of Super-Yang-Mills theory is indeed obtained
from quantum loops in Supergravity, confirming the Maldacena
conjecture to this order.

The metric for $d+1$-dimensional Anti-de Sitter ($AdS_{d+1}¥$)
space can be written \be ds^2=
G_{\mu\nu}\,dX^\mu\,dX^\nu=dr^{2}+z^{{-2}}¥\eta_{ij}¥dx^{i}¥dx^{j}¥\,,
\quad z=\exp(r/l)\,, \label{ads} \ee with $\eta_{ij}$ the
$d$-dimensional Minkowski metric, and $l$ a constant. The Riemann
tensor is \be R_{\mu\nu\lambda\rho}¥=-{1\over
l^2}\left(G_{\mu\lambda}¥G_{\nu\rho}¥
-G_{\nu\lambda}¥G_{\mu\rho}¥\right)\,, \label{max} \ee which
leads to the (d+1)-dimensional Einstein equation \be R_{\mu\nu} =
-{d\over l^2}
 G_{\mu\nu}\,, \label{einsteineqn}\ee
with the cosmological constant $\Lambda = -d(d-1)/2l^2$ and $R =
-d(d+1)/l^{2}$. The boundary of this space occurs at $r=-\infty$
and at $r=\infty$ (which is just a point because the
`warp-factor', $z^{-2}¥$ vanishes there). Following
\cite{Henningson} we will treat this boundary as though it
occurred at $z=\exp (r_0/l)\equiv\tau$ and send the cut-off,
$\tau$, to zero at the end of our calculations, so the metric
restricted to the boundary is $\eta_{ij}/\tau^{2}¥$.

Maldacena
conjectured an equivalence between Type IIB String Theory
compactified on $AdS_{5}¥\times S^{5}¥$ (the bulk theory)
and $4$-dimensional
Super-Yang-Mills theory in Minkowski space (the
boundary of $AdS_{5}$) with gauge-group $SU(N)$. The string
compactification is driven by the presence of $N$ D3-branes
which generate a 5-form flux. Performing the functional integral for
the Superstring theory with the fields taking prescribed values on the
boundary of $AdS_{5}$ is meant to reproduce the generating functional
of Green functions for operators, $\Omega$, in the Yang-Mills theory.
The identification
between the boundary fields and the various $\Omega$ has been made
on the basis of symmetry for many operators, and there is a precise
relationship between the couplings in the two theories.
After a Wick rotation the conjecture may be written as \be \int
{\cal D}\Phi\,e^{-S_{IIB}¥}¥\Big|_{\Phi(r=-\infty)=\hat\Phi}¥
=\int {\cal D} A\,e^{-S_{YM}¥+\int d^{4}¥x\,\hat\Phi\Omega(A)}¥
\,. \label{MC} \ee This is only a formal statement since the
left-hand-side written in terms of the string fields of IIB
Superstring theory is not well-defined.  (Nonetheless, an
observation that will be crucial later is that this ill-defined
functional of the boundary fields, $\Psi[\hat\Phi]$ corresponds to
Feynman's construction of the vacuum wave-functional.) Taking
$\Omega$ as the stress-tensor of the gauge theory allows the
source $\hat\Phi$ to be interpreted as a perturbation to the
Minkowski metric, so that in the absence of other sources the
right-hand-side becomes the partition function for
Super-Yang-Mills theory in the perturbed metric, $g_{ij}¥$, whose
logarithm we denote by $F[g]$. The Weyl anomaly, ${\cal A}$, is
then $\delta F=\int d^4x\,{\sqrt g} \delta\sigma {\cal A}$ when
$\delta g_{ij}=2\delta\sigma g_{ij}$. On general grounds,
\cite{bonora} \cite{Duff1}, ${\cal A}=a \,E+c\,I$ where $E$ is the
Euler density, $(R^{ijkl}R_{ijkl}-4R^{ij}R_{ij}+R^{2})/64$, and
$I$ is the square of the Weyl tensor,
$I=(-R^{ijkl}R_{ijkl}+2R^{ij}R_{ij}-R^{2}/3)/64$. A one-loop
calculation \cite{Duff1} gives ${\cal A}$ as the sum of
contributions from the six scalars, two fermions and gauge vector
of the Super-Yang-Mills theory, (all in the adjoint with dimension
$N^2-1$) \be\label{cftaa}{\cal A}={(6s+2f+g_v)(N^2-1)\over
16\pi^2}\,.\ee When the heat-kernel coefficients $s$, $f$, and
$g_v$ are expressed in terms of $E$ and $I$ this becomes
\be\label{cfta}{\cal A}=-{(N^2-1)(E+I)\over \pi^2},\ee so
$a=c=-(N^2-1)/(2\pi^{2})$ and supersymmetry protects this from
higher-loop corrections. Equation (\ref{MC}) shows how to find
$F[g]$ in the Superstring theory. At leading order in $N$ we can
replace strings by fields, and neglect all the fields in the
resulting Supergravity theory except the graviton, so $S_{IIB}¥$
reduces to the Einstein-Hilbert action with cosmological term,
whilst the functional integral itself can be computed in the
saddle-point approximation and so reduces to the exponential of
minus the action computed with the metric satsfying Einstein's
equation and coinciding (up to a conformal factor) with $g_{ij}¥$
when restricted to the boundary. By solving this boundary value
problem in perturbation theory Henningson and Skenderis
\cite{Henningson} were able to compute ${\cal A}$ to leading order
in large-$N$.

Rather than use perturbation theory the Weyl anomaly can be calculated
more simply by using an exact solution to the Einstein equations that is more general than (\ref{ads}). Replacing the Minkowski metric $\eta_{ij}¥$
by a d-dimensional Einstein metric $\hg$, (${\hat R}_{ij} = {\hat R}\,
\hg_{ij}/4$ and ${\hat
R}=$ constant), and modifying the warp-factor
\begin{equation}
ds^2 = G_{\mu\nu}\,dX^\mu\,dX^\nu=dr^2 + z^{-2}¥\, e^{\rho}
\hg_{ij}(x)\, dx^i dx^j \, ,\quad e^{\rho/2}= 1-C\,z^{2}\,,
\quad¥C={l^2 {\hat R} \over 4\,d(d-1)}\,,\label{ads1}
\end{equation}
results in a bulk metric that still satisfies (\ref{max})
and (\ref{einsteineqn}).
The metric restricted to the `boundary' at $z=\tau=$ is $\hg$ up
to a conformal factor. We now specialise to $d=4$. The greater
generality of this metric is useful because it allows us to
calculate the anomaly coefficients $a$ and $c$ by making special
choices for $\hg$. Were we to calculate the anomaly for Ricci flat
$\hg$, so that $E=-I=R^{ijkl}¥R_{ijkl}¥/64$ we would find the
combination $a-c$. By taking instead a $\hg$ for which $\hat
R_{ijkl}¥=\hat R¥\left(\hg_{ik}¥\hg_{jl}¥
-\hg_{jk}¥\hg_{il}¥\right)/12$ so that $I=0$ and $E=\hat
R^{2}¥/384$ we would obtain the coefficient $a$. The
Einstein-Hilbert action evaluated in this metric is \be S_{\rm
EH}={1\over 16\pi G_N}\int d^5 X{\sqrt G}\left(R+2\Lambda\right)
=-{1\over 2\pi G_N l^2}\int {dr\,d^4 x\over z^4}\sqrt{\hg}
\,(1-Cz^2)^4 \ee

The integral over $r$ diverges as the boundary is approached,
hence the need for the cut-off. Because the anomaly depends on
just two numbers, $a$ and $c$, it is sufficient to consider a Weyl
scaling of the boundary metric, $\delta g_{ij}=2\delta\sigma
g_{ij}$, with constant $\delta\sigma$, and this can be achieved by
keeping $\hg$ fixed, but varying $r_0$ by $\delta
r_0=l\delta\sigma$, so \be \delta F\approx -\delta\sigma\,l{\pd
\over \pd r_0}S_{\rm EH}= -{\delta\sigma\over 2\pi G_N l}\int {d^4
x\over \tau^4}\sqrt{\hg} \,(1-C\tau^2)^4 \ee The divergent parts
of this as $\tau\downarrow 0$ can be cancelled by adding
counter-terms to the action, but the finite contribution,
proportional to $C^2$, cannot, so we obtain the bulk tree-level
contribution to the anomaly as \be {\cal A}_{\rm tree}=-{3l^3\over
48^2\pi G_N}R^2=-{l^3\over 2\pi G_N}(E+I)\,. \ee The gravitational
coupling is related to $N$ and $l$ via $G_N=\pi l^3/ (2N^2)$,
\cite{Henningson}, so this reproduces the leading term in
(\ref{cfta}). (Note that it is easy to check that the
Gibbons-Hawking boundary action does not contribute to this
calculation of $\cal A$ so we have not included it in our
discussion).

To go beyond the leading order and compute the bulk one-loop
contribution to the Weyl anomaly, $\delta{\cal A}$,
 requires making sense of the left-hand-side
of (\ref{MC}). We will approximate the IIB Superstring theory by
IIB Supergravity, which means neglecting higher orders in
$\alpha'$. Even so the left-hand-side is ill-defined. At the time
we began our work there was no known action for this theory, but
rather a set of classical equations of motion consistent with
Supersymmetry which were analysed in \cite{Kim} to obtain the mass
spectrum for the theory compactified on $AdS_{5}¥\times S^{5}¥$.
It was not obvious that these equations of motion could be derived
from an action, but in Section 6 we construct one that yields the
equations of motion to quadratic order in the quantum fluctuations
of the fields which is what is needed to  obtain the one-loop
contribution to the anomaly in the bulk theory. Integrating out
these fluctuations would give a functional determinant for each of
the infinite number of fields in the compactification. To compute
these in the conventional fashion requires the use of the
heat-kernels for differential operators defined on a
five-dimensional manifold with boundary, again these were unknown
at the time we began. We will adopt a different approach based on
the interpretation of $\Psi$ as a wave-functional which satisfies
a functional Schr\"odinger equation from which it can be
constructed. Because this is a Hamiltonian approach it involves
four-dimensional differential operators whose heat-kernel
coefficients are already tabulated. Furthermore, because it treats
the five-dimensional bulk fields in terms of their values on the
boundary it is ideally suited to discussing the Maldacena
conjecture.

Consider, for the sake of illustration, a scalar field of mass $m$.
To quadratic order in the field there are no interactions other than those
with the background metric, so the action is
\bq
S_\phi&=&{1\over 2}\int d^5 X{\sqrt G}\left(G^{\mu\nu}\partial_\mu\phi\,
\partial_\nu\phi+m^2\phi^2\right)\nonumber \\
&=&{1\over 2}\int {d^4x\,dr\over
z^4}{\sqrt\hg}\,e^{2\rho}\,\left(\dot\phi^2
+z^2 e^{-\rho}\hg^{ij}\partial_i\phi\,\partial_j\phi+m^2\phi^2\right)\nonumber \\
&\equiv&{1\over 2}\int {d^4x\,dr\over
z^4}{\sqrt\hg}\,e^{2\rho}\,\phi\,\Omega_s\,\phi \,,\label{sca} \eq
(here the dot denotes differentiation with respect to $r$) whilst
the norm on fluctuations of the field, from which the functional
integral volume element ${\cal D}\phi$ can be constructed is \be
||\delta\phi||^2=\int d^5 X{\sqrt G}\,\delta\phi^2 =\int
{d^4x\,dr\over z^4}{\sqrt\hg}\,e^{2\rho}\,\delta\phi^2\,. \ee We
will interpret the co-ordinate $r$ as Euclidean time, so to construct a \Sc equation we first re-define the field by setting
$\phi=z^2\,e^{-\rho}\varphi$ to make the `kinetic' term in the
action into the standard form, and remove the explicit
$r$-dependence from the integrand of the norm. The action becomes
\bq &&S_\phi\nonumber\\
&=&{1\over 2}\int
{d^4x\,dr}{\sqrt\hg}\,\left(\dot\varphi^2+z^2 e^{-\rho}\varphi
\left(\Box+{{\hat R}\over 6}\right)\varphi+ \left(m^2+{4\over
l^2}\right)\varphi^2\right)-{1\over 2}\int d^4x\,{\sqrt\hg}
\left(\dot\rho+{2\over l}\right)\varphi^2\,,\nonumber\\
&=&S_\varphi+S_b\,,
\eq
where $\Box$ is the 4-dimensional covariant Laplacian
constructed from $\hg$. Note that
$\Box+{{\hat R}\over 6}$ is the operator associated with a
conformally coupled four-dimensional scalar field, its appearance should not be a
surprise given that isometries of $AdS$ act as conformal transformations on
its boundary. Also note that the mass has been
modified to an effective mass $\sqrt{m^2+4/l^2}$.

The one-loop contribution of a scalar field to the left-hand-side
of (\ref{MC}) is \be \int {\cal
D}\phi\,e^{-S_{\phi}¥}¥\Big|_{\phi(r=r_0)=\hat\phi}¥ \ee where
the `boundary' is taken at the cut-off, $r_0$, rather than
$r=-\infty$. In terms of the `canonical' field $\varphi$ this is
\be e^{-S_b}\int {\cal
D}\varphi\,e^{-S_{\varphi}¥}¥\Big|_{\varphi(r=r_0)=\hat\varphi}
¥\equiv e^{-S_b+W[\hat\varphi,g]} \, \ee Since the integral is
Gaussian, $W$ takes the form \be W[\hat\varphi]=F+{1\over 2}\int
d^4 x\,\sqrt{\hg}\,\hat\varphi\Gamma\,\hat\varphi \ee where
$\Gamma$ is a differential operator and $F=-{1\over 2}\log {\rm
Det} \, \Omega_s$ is the free-energy of the scalar field, whose
variation under a Weyl transformation is the goal of our
computation. As we have already observed, this functional integral
can be interpreted as the vacuum wave-functional at Euclidean
`time' $r_0$, and so it satisfies a functional \Sc equation that
can be read-off from the action $S_\varphi$, thus if $\Psi=\exp
W[\varphi]$ then \be {\pd \over \pd r_0}\,\Psi=-{1\over 2} \int
d^4x \sqrt{\hg}\, \Big\{-\hg^{-1}\, {\dl^2 \over \dl\varphi^2} +
\tau^2 e^{-\rho}\varphi \left(\Box+{{\hat R}\over
6}\right)\varphi+\left(m^2+{4\over l^2}\right) \varphi^2
\Big\}\,\Psi\,. \label{sch2} \ee So $\Gamma$ satisfies \be {\pd
\over \pd r_0}\Gamma=\Gamma^2-\tau^2 e^{-\rho}\left(\Box+{{\hat
R}\over 6}\right)-\left(m^2+{4\over l^2}\right)\,,\label{gam} \ee
which can be solved in powers of the differential operator by
expanding \be \Gamma=\sum_{n=0}^\infty b_n(r_0) \left(\Box+\hat
R/6\right)^n\,.\label{ex} \ee This gives \be
b_0=\pm\sqrt{m^2+{4\over l^2}} \ee in which we take the minus sign
to give a normalisable wave-functional. The other coefficients in
(\ref{ex}) have the property that we will make use of later of
vanishing as the cut-off, $r_0$ is taken to $-\infty$.

The free-energy is determined in terms of the functional trace of
$\Gamma$: \be {\pd \over \pd r_0}F={1\over 2}{\rm Tr}\, \Gamma \ee
which can be regulated using a heat-kernel \be {\rm Tr}\,
\Gamma=\sum_{n=0}^\infty b_n(r_0)\,\left(-{\pd \over \pd s}
\right)^n {\rm Tr}\,\exp \left(-s\left(\Box+\hat R/6\right)\right)
\ee with $s$ small. The heat-kernel has the well-known Seeley-de
Witt expansion for small $s$ \be {\rm Tr}\,\exp
\left(-s\left(\Box+\hat R/6\right)\right) =\int d^4
x\,{\sqrt\hg}{1\over 16\pi^2s^2}\left(1+s\,a_1(x)
+s^2\,a_2(x)+s^3\,a_3(x)+..\right)\label{sdw} \ee with $\sqrt\hg
\,a_2=\sqrt g \,8(2E-3I)/45$. As $s$ is made smaller and $|r_0|$
larger the only surviving contribution comes from $1$, $a_1$ and
$a_2$. The coefficients of $1$ and $a_1$ diverge, but that of
$a_2$ is finite.

As well as fixing $W$, these equations directly determine the Weyl
anomaly\footnote{In our calculation of the anomaly, to remove the
cutoff dependence from the functional inner-product we imposed
non-standard boundary conditions on the bulk field. The resulting
wave-functional differs only by a boundary term from that obtained
by the standard procedure (diagonalising the asymptotic part of
the bulk field corresponding to the larger scaling dimension) and
gives the same Weyl anomaly. For masses in a certain range, it is
also possible to diagonalise the asymptotic corresponding to the
smaller scaling dimension \cite{witten}; this gives a different
result for the anomaly. For certain compactifications, for example
when $S^5$ is replaced by $T^{1,1}$, this ambiguity becomes
important, but in the present case the spectrum contains no modes
with masses in the appropriate range. All of this is discussed in
more detail in \cite{d}.}. As in the leading order calculation we
consider a constant scaling of the boundary metric resulting from
a shift in $r_0$, $\delta r_0=l\delta\sigma$, so $ \int
d^4x\,\sqrt g\,\delta{\cal A}=l{\pd \over \pd r_0}F={l\over 2}{\rm
Tr}\, \Gamma $. The divergent parts of this can be cancelled by
adding counter-terms to $F$, but the finite contribution
proportional to $a_2$ cannot, so we obtain the anomaly as $
\delta{\cal A}=-{\sqrt{l^2m^2+4}}\,a_2/(32\pi^2) $. Now the mass
dependence can be neatly expressed in terms of the scaling
dimension of the field restricted to the boundary, $\Delta$,
because $\sqrt{l^2m^2+4}=\Delta-2$ so we arrive at \be \delta{\cal
A}=-{\Delta-2\over 32\pi^2}\,a_2\,.\label{basicr} \ee Although we
have derived this formula for a scalar field it applies to all the
species of fields in IIB Supergravity. To see this requires
decomposing the appropriate action into `canonical' fields so as
to identify the appropriate four-dimensional operators and
effective masses. We will describe the details of this in the
subsequent sections of this paper, but the upshot of this
decomposition of five-dimensional fields into four-dimensional
variables is to introduce into the four-dimensional operators
precisely those couplings to $\hat R$ that render them conformally
covariant. Thus $a_2$ for a five-dimensional gauge field is the
combination of heat-kernel coefficients for the operator
associated with a four-dimensional gauge field, just as that for a
minimally coupled five-dimensional scalar is associated with a
conformally coupled four-dimensional scalar.

The scaling dimensions $\Delta$ are related to the bulk masses
which were originally worked out in \cite{Kim}. In Table 1 we
display the corresponding values of $\Delta-2$. The multiplets are
labelled by an integer $p\ge 2$, and the fields form
representations of $SU(4)\sim SO(6)$. The four-dimensional
heat-kernel coefficients have also been known for a long time and
we use the values given by \cite{Duff2,Barv}. In Table 2 we list
these for the cases of a Ricci flat boundary and for a boundary
of constant $\hat R$.

If we denote the values of $a_{2}$ for the fields $\phi$, $\psi$,
$A_\mu$, $A_{\mu\nu}$, $\psi_\mu$, $h_{\mu\nu}$ by $ s,f,v,a,r, $
and $g$ respectively then the contribution from a generic ($p\ge
4$ ) multiplet is \bq \left(\sum (\Delta-2)a_2\right)_{p\ge 4}&=&
(-4s+4a+r+f+2v){p\over 3}\nonumber\\
&&-(105s+g+26a+8r+72f+48v){p^{3}\over 12}\nonumber\\
&&+(16v+20f+10a+4r+25s+g){p^{5}\over 12} \eq whilst for the $p=3$
multiplet it is \be \left(\sum (\Delta-2)a_2\right)_{p=3}=
244f+18g+266s+218v+148a+64r\,. \ee The $p=2$ multiplet contains
gauge fields requiring the introduction of Faddeev-Popov ghosts.
Their parameters are given in Table 3 along with the decomposition
of the five-dimensional components of fields into four-dimensional
pieces. \be 12 v- 30  s +6 r-10f +2g \ee and if we include the
scalars, spinors and antisymmetric tensors the total contribution
of the $p=2$ multiplet is \be \left(\sum
(\Delta-2)a_2\right)_{p=2}=12 v- 6  s +6 r+6f +2g +12a \ee
Substituting the values of the heat kernel coefficients for a
Ricci flat boundary shows that the contribution of each
supermultiplet vanishes implying that $a=c$ \cite{testn}. However
if we do not specialise to this case we have to deal with the sum
over multiplets labelled by $p$. We will evaluate this divergent
sum by weighting the contribution of each supermultiplet by $z^p$.
The sum can be performed for $|z|<1$, and we take the result to be
a regularisation of the weighted sum for all values of $z$.
Multiplying this by $1/(z-1)$ and integrating around the pole at
$z=1$ gives a regularisation of the original divergent
sum\footnote{This regularisation is equivalent to simply taking a
cut-off $p=\Lambda$ in the summation of supermultiplets with
$p\geq4$ and removing $\Lambda$ dependent divergent terms at
$\Lambda \rightarrow \infty$ from the regularised sum. Both these
regularisations preserve supersymmetry, as they must.}. In this
way all the members of any given supermultiplet are treated on an
equal footing. This yields \be \sum (\Delta-2)a_2=8s+4f+2v \ee
which remarkably depends only on the heat-kernel coefficients of
fields in the Super-Yang-Mills theory. In the next section we will
see that decomposing a five-dimensional vector into longitudinal
and transverse pieces and solving the Schr\"odinger equation for
them relates the heat-kernel coefficient for a vector field, $v$,
to that for a four-dimensional gauge-fixed Maxwell field, $v_0$,
by $v = v_0 + 2s - 2s_0$. $s_0$ is the coefficient for a minimally
coupled four-dimensional scalar (Faddeev-Popov ghost), showing
that $v-2s = v_0 -2s_0 = g_v$ \cite{mnu-prep}. Therefore we
finally arrive at the one-loop contribution to the Weyl anomaly
\be \delta {\cal A}=-\sum {(\Delta-2)a_{2}\over
32\pi^{2}}=-{6s+2f+g_v\over 16\pi^2} \label{afsum} \ee which is
precisely what is needed to reproduce the subleading term in the
exact Weyl anomaly of Super-Yang-Mills theory and verify the
Maldacena conjecture.

A final point concerns the finiteness of the boundary theory. The
divergence of the coefficients $a_0$ and $a_1$ in (\ref{sdw})
renormalises the boundary cosmological and Newton constants,
respectively, but we would expect these renormalisations to
disappear in the full theory. If we wrote down (\ref{sdw}) in some
superfield formalism, we would have to take the same proper-time
separation for fields of different spin. So it makes sense to sum
the contributions of all Supergravity fields to these
coefficients. If we do so, the total $a_0$ contribution cancels by
virtue of the equal number of bosonic and fermionic modes. The
total $a_1$ contribution also cancels, but only after we apply the
same regularisation that we used to sum the $a_2$ coefficients. So
we find as expected that there is no overall renormalisation of
the boundary Newton or cosmological constants \cite{finiteness}.
\section{Weyl Anomaly for Fermions}

The Euclidean action for a spin-1/2 fermion in the metric
(\ref{ads1}) is

\be \int d^{d+1}x\sqrt G\bar\psi(\gamma^\mu D_\mu-m)\psi. \ee

The spin-covariant derivative is defined via the funfbein

\be V^\alpha_0={1\over z}\delta^\alpha_0,\qquad V^\alpha_i={1\over
z}e^{\rho/2}\tilde V^\alpha_i, \ee

where $\tilde V^\alpha_i$ is the vierbein for the boundary metric.
Making the change of variables $\psi=z^2e^{-\rho}\tilde\psi$
causes the volume element in the path-integral to become the usual
flat-space one, and the kinetic term in the action acquires the
usual form. The action can be written

\be \int d^{d+1}x \bar{\tilde\psi}
\left(\gamma^0\partial_0+ze^{-\rho/2}\gamma^i\tilde
D_i-m\right)\tilde\psi. \ee

The $D_i$ derivative is spin-covariant with respect to the
boundary metric.

We impose the following boundary conditions on $\tilde\psi$:

\be Q_+\tilde\psi(0,x)=u(x)=Q_+u(x),\qquad\tilde\psi^\dagger(0,x)
Q_-=u^\dagger(x)=u^\dagger(x)Q_-, \ee

for some local projection operators $Q_\pm$. The remaining
projections are represented by functional differentiation. The
partition function takes the form

\be \Psi[u,u^\dagger]=\exp[f+u^\dagger\Gamma u], \label{fw}\ee and
the Schr\"odinger equation that it satisfies can be written

\be {\partial\over\partial r_0}\Psi=-\int d^dx\left(u^\dagger
Q_-+{\delta\over\delta u}Q_+\right)h\left(Q_+
u+Q_-{\delta\over\delta u^\dagger}\right)\Psi, \label{fa}\ee

where $h=\tau e^{-\rho/2}\gamma^0\gamma^i\tilde D_i-\gamma^0m$.
Assume without loss of generality that $m\ge0$. If we make the
specific choice $Q_\pm=\hf(1\pm\gamma^0)$, we can write (\ref{fa})
as

\be {\partial\over\partial
r_0}\Psi=-\left[mu^\dagger{\delta\over\delta
u^\dagger}-m{\delta\over\delta u}u-\tau
e^{-\rho/2}u^\dagger\gamma\cdot\tilde Du+\tau
e^{-\rho/2}{\delta\over\delta u}\gamma\cdot\tilde
D{\delta\over\delta u^\dagger}\right]\Psi. \ee

Acting on (\ref{fw}) this implies that

\be \dot\Gamma=-2m\Gamma+\tau e^{-\rho/2}\gamma\cdot\tilde
D-\Gamma^2\tau e^{-\rho/2}\gamma\cdot\tilde D, \label{feq}\ee

while $f$ satisfies

\be \dot f=\hf{\rm Tr}(-m+\Gamma\tau e^{-\rho/2}\gamma\cdot\tilde
D).\label{ffe} \ee

The factor of 1/2 takes into account the fact that the trace is
over constrained variables. So that we can regulate this with a
heat-kernel, we expand $\Gamma$ in terms of the positive-definite
operator $(\gamma\cdot\tilde D)^2$:

\be \Gamma=\gamma\cdot\tilde D\sum_{n=0}^\infty
d_n(r_0)(\gamma\cdot\tilde D)^{2n}. \ee

Notice that the coefficients $d_n$ {\em all} vanish as
$r_0\--\infty$. The equation (\ref{feq}) is easily solved in terms
of Bessel functions, but to regulate (\ref{ffe}) we again use a
heat-kernel expansion

\be {\rm Tr}(-m+\Gamma\tau\gamma\cdot\tilde
D)=\left(\sum_{n=0}^\infty d_n(r_0)\left(-{\partial\over\partial
s}\right)^{n+1}-m\right){\rm Tr}\exp\left(-s(\gamma\cdot\tilde
D)^2\right), \ee

where the heat-kernel has a Seeley-de Witt expansion like
(\ref{sdw}). The contribution proportional to the $a_2$
coefficient of $(\gamma\cdot\tilde D)^2$ is finite as $s\-0$ and
$r_0\--\infty$ and determines the anomaly, which is therefore
proportional to $m$. But since $m=\Delta-2$ we have as before

\be \delta{\cal A}=-{\Delta-2\over 32\pi^2}\,a_2\,. \ee

For the spin-3/2 Rarita-Schwinger field the action that we obtain
by diagonalising the five-dimensional action has the same form as
the spin-1/2 field, and the Schr\"odinger equation takes the form
(\ref{fa}). Thus the above discussion is essentially unchanged,
leading again to an anomaly proportional to $m=\Delta-2$.

\section{Vector Fields}

We now wish to demonstrate that the result (\ref{basicr}) extends
to higher spin fields as well. We begin by considering the
decomposition of a five-dimensional vector gauge field in $AdS$ in
terms of `canonical' fields from which we can construct a
functional \Sc equation. The classical action for a $U(1)$ gauge
field in the metric (\ref{ads1}) is
\begin{equation}
S_{\rm gv}={1\over 2}\int d^{4}¥x\,dr\,{\sqrt\hg}\left((\dot A_{i}¥-\partial_{i
}¥A_{r}¥)(\dot A_{j}¥-\partial_{j}¥A_{r}¥)\,\hg^{ij}¥e^{\sigma}¥+
(\partial_{i}¥A_{j}¥-\partial_{j}¥A_{i}¥)(\partial_{r}¥A_{s}¥-\partial_{s}
¥A_{r}¥)
\hg^{ir}¥\hg^{js}¥\right)\, , \label{gvaction}
\end{equation}
where we have set $\exp\sigma=\exp (\rho)/z^2$. We choose as a
gauge condition that $A_{r}¥$ be constant, so that the
Euler-Lagrange equation corresponding to varying $A_{r}¥$,
\begin{equation}
    \nabla_{i}¥\,(\hg^{ij}¥\dot A_{j}¥)=0
     \label{gcondition}
 \end{equation}
must be imposed as a constraint, with $\nabla$ the Levi-Civita
connection constructed from $\hg_{ij}$. If we change variables
from $A$ to $\ca=e^{\sigma/2}¥A$ the kinetic term assumes the
canonical form and the action becomes

\begin{equation}
S_{\rm gv}={1\over 2}\int d^{4}¥x\,dr\,{\sqrt\hg}\left(\dot \ca_{i}\dot
\ca_{j}\,\hg^{ij}+{1\over l^2}\ca_{i}
\ca_{j}\,\hg^{ij}+
e^{-\sigma}¥(\partial_{i}¥\ca_{j}¥-\partial_{j}¥\ca_{i}¥)(\partial_{r}
¥\ca_{s}¥-
\partial_{s}¥\ca_{r}¥)\,
\hg^{ir}¥\hg^{js}¥\right)\, , \label{gvaction2}
\end{equation}
whilst the general co-ordinate invariant inner product on variations
of the gauge field, (from which the functional integral volume element
can be constructed) also takes the form appropriate to a canonical
field theory

\begin{equation}
    ||\,\delta A \,||^{2}¥=\int d^{4}¥x\,dr{\sqrt\hg}\,e^{\sigma}¥\,
    \delta A_{i}¥
    \delta A_{j}¥\,\hg^{ij}=\int d^{4}¥x\,dr{\sqrt\hg}\,\delta
    \ca_{i}¥
    \delta \ca_{j}¥\,\hg^{ij}\, . \label{gvinnerprod}
\ee
We can now write down the functional \Sc equation
satisfied by
\be
\Psi=\int {\cal D}A \,e^{-S}\Big |_{A(r=r_0)=\hat A}
\ee
Treating $r$ as
a Euclidean time and quantising using $\dot\ca_{i}¥\rightarrow
-(\hg_{ij}¥/{\sqrt \hg})\,
{\delta}/\delta\ca_{j}¥$
\bq
&&
{\partial\over\partial r_0}\Psi=\nonumber\\
&&-{1\over 2}\int d^{4}¥x\,{\sqrt\hg}\Big(
-{1\over \hg}\hg_{ij}{\delta^{2}¥\over\delta\ca_{i}\delta \ca_{j}}+{1\over l^2}\ca_{i}
\ca_{j}\,\hg^{ij}+\nonumber\\
&&e^{-\sigma}¥(\partial_{i}¥\ca_{j}¥-\partial_{j}¥\ca_{i}¥)(\partial_{r}¥\ca_{s}¥-
\partial_{s}¥\ca_{r}¥)\,
\hg^{ir}¥\hg^{js}¥\Big)\,\Psi\, . \label{gvsch1}
\eq
The constraint is imposed weakly, i.e. as a condition on
$\Psi$:
\be
\nabla^{i}¥\left(g_{ij}¥{\delta\over\delta\ca_{j}¥}-{\dot\sigma\over 2}\ca_{i}¥
\right)\Psi=0\,.
\ee
This can be analysed by decomposing $\ca$ as
\be
\ca=\tilde\ca+\nabla\varphi\,,\quad \nabla_{i}¥ (\hg^{ij}¥\tilde\ca_{j}¥)=0\,,
\ee
so that
\be
{\delta\over\delta\ca_{j}¥}={\delta\over\delta\tilde\ca_{j}¥}+\nabla^{j}¥
\Box^{-1}¥{\delta\over\delta\varphi¥}
\ee
The constraint becomes
\be
\left({1\over \sqrt \hg}{\delta\over\delta\varphi¥}+{\dot\sigma\over 2}
\Box\varphi
\right)\,\Psi=0\,,
\ee
with solution
\be
\Psi=\exp\left(-{\dot\sigma\over 4}\int d^{4}¥x\,{\sqrt\hg}\varphi\,
\Box\,\varphi\right)\,\Psi_{0}¥[\tilde\ca]
\ee
(\ref{gvsch1}) now implies that $\Psi_0$ satisfies
\bq
&&
{\partial\over\partial r}\Psi_{0}¥=\label{gvsch}\\
&&-\left({1\over 2}\int d^{4}¥x\,dr\,{\sqrt\hg}\left( -{1\over
\hg}\hg_{ij}{\delta^{2}¥\over\delta\tilde\ca_{i}\delta
\tilde\ca_{j}}+{1\over l^2}\tilde\ca_{i} \tilde\ca_{j}\,\hg^{ij}+
e^{-\sigma}¥\,\tilde\ca_{i}¥\left(\Box+{\hat R\over
4}\right)\tilde\ca_{j}¥ \hg^{ij}\right)+{\dot\sigma\over 4}{\rm
Tr} 1_{s}¥ \right)\,\Psi_{0}¥\, \nonumber.  \eq Comparing this
to (\ref{sch2}) shows that the Weyl anomaly of the gauge field is
given by an effective mass of $1/l^2$ and the heat-kernel of the
operator $\Box+{\hat R\over 4}$ acting on divergenceless
four-dimensional vectors. Now the projector onto divergenceless
four-dimensional vectors is $ {\cal
P}^i_j=\delta^i_j+\nabla_j\Box^{-1}\nabla^i $ and $(\Box+\hat
R/4)\nabla^j=\nabla^j\Box$ when acting on scalars, so \be
\left(\Box+\hat R/4\right)^n{\cal P}= \left(\Box+\hat
R/4\right)^n+\nabla \Box^{n-1}\nabla \ee hence \be {\rm
Tr}\,\left(e^{-s(\Box+\hat R/4)}{\cal P}\right)= {\rm
Tr}\,e^{-s(\Box+\hat R/4)}-{\rm Tr}\,e^{-s\Box} \ee i.e. the trace
of the heat-kernel for $\Box+\hat R/4$ acting on divergenceless
vectors is equal to the trace of the heat-kernel for $\Box+\hat
R/4$ acting on unconstrained four-dimensional vectors minus the
trace of the heat-kernel for the operator $\Box$ acting on
scalars,\cite{strong}. Given the origin of the term
${\dot\sigma\over 4}{\rm Tr} 1_{s}¥$ we regulate it using the
heat-kernel for $\Box$ acting on scalars. If we denote the $a_2$
Seeley-de Witt coefficients for the $\Box+{\hat R\over 4}$ acting
on unconstrained four-dimensional vectors by $v_0$ and for $\Box$
acting on scalars by $s_0$, then the 1-loop contribution of the
gauge-field to the boundary Weyl anomaly is $\delta{\cal
A}=-(v_0-2s_0)/(32\pi^2)$. This combination of heat-kernel
coefficients is precisely that that arises in the Weyl anomaly of
a four-dimensional gauge-field after gauge-fixing in the
Lorentz-gauge, the $2s_0$ corresponding to the Faddeev-Popov
ghosts which are minimally coupled to the metric. This should not
be surprising since the two pieces of this expression do not
correspond to conformally invariant actions, but their sum, a
four-dimensional gauge-theory, does.

The $S^5$ compactification of IIB Supergravity produces
mass terms to be added to the action $S_{\rm gv}$ of the form
\be
S_{\rm mgv}={m^2\over 2}\int d^{4}¥x\,dr\,{\sqrt G}\,G^{\mu\nu}A_\mu
A_\nu={m^2\over 2}\int d^{4}¥x\,dr\,{\sqrt\hg}\left(e^{2\sigma
}A_r^2
+e^\sigma \hg^{ij}A_iA_j\right)\,.
\ee
This breaks the gauge invariance and couples $A_r$ to the longitudinal
part of $A_i$. To decouple these degrees of freedom we change
variables to $\tilde A_i$, which is constrained to be
divergenceless, and $u$ and $w$, where
\be
A_i=\tilde A_i+\nabla_i\left(u+\Box^{-1/2}e^{-\sigma}\partial_r
\left(e^{3\sigma/2}w\right)\right),\quad A_r=\partial_r u+\Box^{1/2}
e^{-\sigma/2}w\,,
\ee
so that the mass term becomes
\be
S_{\rm mgv}={m^2\over 2}\int d^{4}¥x\,dr\,{\sqrt\hg}e^{2\sigma
}\left(e^{-\sigma} \hg^{ij}\tilde A_i\tilde A_j+\dot u^2+e^{-\sigma}u\,
\Box \,u +w\,\Omega_w \,w\right)\,,
\ee
where
\be
\Omega_w\,w=e^{-\sigma/2}\partial_r\left(e^{-\sigma}\partial_r
\left(e^{3\sigma/2}w\right)\right)-e^{-\sigma}\Box w
\ee
$u$ decouples from $S_{gv}$ which becomes
\begin{equation}
S_{\rm gv}={1\over 2}\int d^{4}¥x\,dr\,{\sqrt\hg}\left(
\partial_r \tilde A_{i}\,\partial_r\tilde A_{j}\,\hg^{ij}¥e^{\sigma}
+\tilde A_i\left(\Box+\hat R/4\right)\,\tilde A_j\hg^{ij}
+e^{2\sigma}\,w\,\Omega_v^2 \,w\right)
\ee
The norm on fluctuations of the field copies the form of the
mass term:
\begin{equation}
    ||\,\delta A \,||^{2}¥=\int d^{4}¥x\,dr\,{\sqrt\hg}e^{2\sigma
}\left(e^{-\sigma} \hg^{ij}\delta\tilde A_i\delta\tilde A_j+
 \delta \dot u^2+e^{-\sigma}\delta u\,\Box \,\delta u +\delta w\,\Omega
 \,\delta w\right)\,, \ee so that the functional integration volume
element factorises into \be {\cal D} A={\cal D} \tilde A\,{\cal D}
u\,\sqrt{{\rm Det}\,\Omega_s}\,{\cal D} w\,\sqrt{{\rm
Det}\,\Omega_w}
\ee where $\Omega_s$ is the same operator that
occurred earlier in the discussion of the scalar field
(\ref{sca}). When $m$ is non-zero the integral over $u$ in
$\int{\cal D}A\,\exp(-S_{\rm gv}-S_{\rm mgv})$ generates
$1/\sqrt{{\rm Det}\,\Omega_s}$ which cancels the corresponding
Jacobian factor, and the integral over $w$ generates $1/\sqrt{{\rm
Det}\,\Omega_w(\Omega_w+m^2)}$, part of which cancels the Jacobian
factor for $w$, leaving $1/\sqrt{{\rm Det}\, (\Omega_w+m^2)}$.
Representing this determinant as another functional integral means
that we can re-write the original functional integral as \bq
&&\int{\cal D}A\,e^{-S_{\rm gv}-S_{\rm mgv}}\nonumber\\
&&=\int{\cal D}\tilde A\,e^{-{1\over 2}\int d^{4}¥x\,dr\,{\sqrt\hg}\left(
\partial_r \tilde A_{i}\,\partial_r\tilde A_{j}\,\hg^{ij}¥e^{\sigma}
+\tilde A_i\left(\Box+\hat R/4\right)\,\tilde A_j\hg^{ij}\right)}\nonumber\\
&&\times\int{\cal D} w\,e^{-{1\over 2}\int d^{4}¥x\,dr\,{\sqrt\hg}
\,e^{2\sigma}w\,(\Omega+m^2)\,w}\label{prodf}
\eq
After a partial integration the action in the $w$-integral reduces to that
of a scalar field in $AdS_5$
\bq
&&\int d^{4}¥x\,dr\,{\sqrt\hg}
\,e^{2\sigma}w\,(\Omega+m^2)\,w\nonumber\\
&&=\int d^{4}¥x\,dr\,{\sqrt\hg}
\,\left(e^{-\sigma}\left(\partial_r\left(e^{3\sigma/2}w\right)
\right)^2+e^{\sigma}w\,\Box\,w+e^{2\sigma}m^2w^2\right)\nonumber\\
&&=\int d^{4}¥x\,dr\,{\sqrt\hg} \,e^{2\sigma}\,\left(\dot
w^2+\left(m^2-{3\over 4}
\left(\ddot\sigma+\dot\sigma^2\right)\right)w^2+e^{-\sigma}w\,\Box\,w\right)\,.
\eq Given that $\ddot\sigma+\dot\sigma^2=4/l^2$ we see that the
squared mass has been shifted $m^2\rightarrow m^2-3$. As we saw in
(\ref{basicr}) this contributes to the Weyl anomaly with a
coefficient $\sqrt{(l^2m^2-3)+4}=\sqrt{l^2m^2+1}$. If we make a
change of variables similar to that for a gauge-vector
$\tilde\ca=e^{\sigma/2}\tilde A$ to turn the `kinetic term' into
canonical form then, just as for the gauge-vector the action
acquires a term $\ca^2$ which shifts the mass $m^2\rightarrow
m^2+1/l^2$ so that the coefficient of the Weyl anomaly is again
$\sqrt{l^2m^2+1}$. Thus $\delta{\cal A}$ for the product of
functional integrals in (\ref{prodf}) is $-\sqrt{l^2m^2+1}\,v/
(32\pi^2)$. It remains to identify the combination of heat-kernel
coefficients in $v$. We can do this by comparing this result with
the corresponding calculation for the gauge-vector. When $m=0$ the
functional integral over $u$ does not produce $1/\sqrt{{\rm
Det}\Omega_s}$ because the appropriate part of the action
vanishes, so the Jacobian factor does not cancel, but rather the
integration generates the volume of gauge transformations which
has to be divided out in order to restrict the integral to
physical degrees of freedom. Now the uncancelled Jacobian factor
involves the determinant of the operator associated with a
massless scalar field in $AdS$, so it contributes mith an
`effective mass' $\sqrt 4$ and heat kernel coefficient $s$
(belonging to the conformally coupled operator $\Box-\hat R/6$ in
four-dimensions) and the `wrong' sign, because it is a Jacobian.
So, computed this way the Weyl anomaly of a gauge-vector is
$-(v-2s)/(32\pi^2)$, but since we have already found this to be
$-(v_0-2s_0)/(32\pi^2)$ we conclude that $v=v_0+2s-2s_0$, as
stated in the introduction.
\par

Finally we give an alternative derivation of the above results
using a five-dimensionally covariant formulation, which is useful
in the study of more complicated systems like the graviton case in
Section 4 and the case of an anti-symmetric tensor in Section 5.
The five-dimensional Lagrangian is
\begin{equation}
{\cal L}_A = \sqrt{g}\, [\, {1 \over 4}\, F_{\mu\nu}F^{\mu\nu} +
\hf \, m^2 A_\mu A^\mu \,] \ .
\end{equation}
With $R_{\mu\nu} = -4 l^{-2} g_{\mu\nu}$, the equation of motion
for the massive case can be written as
\begin{equation}
(-\Box - 4 l^{-2} + m^2)\, A_\mu = 0 \ , \qquad \nabla^\mu A_\mu =
0 \ , \label{eom}
\end{equation}
suggesting that we decompose the path-integral variable as $A_\mu
= \hA_\mu + \pd_\mu \varphi$ with $\nabla^\mu \hA_\mu = 0$,
\begin{equation}
Z_A = \int {\cal D}A \, e^{-\int {\cal L}_A} = \int \cD \hA \,
e^{-\int \sqrt{g}[{1 \over 4}F^2(\hA) + \hf m^2 \hA^2]}\, \int \cD
\varphi |\triangle_s|^{\hf} \, e^{-\hf m^2 \int \sqrt{g}\, \varphi
\triangle_s \varphi} \ ,
\end{equation}
where $\triangle_s = -\Box$ acting on the scalar $\varphi$ and the
factor $|\triangle_s|^{\hf}={\rm Det}(-\Box)^{\hf}$ arises from
the Jacobian. For $m^2\not= 0$, the Jacobian is suppressed by the
path-integral for $\varphi$ and we have
\begin{equation}
Z_{massive} = Z_{\hA} = \int \cD \hA \, e^{- \hf \int \sqrt{g}\,
\hA_a (-\Box - 4 l^{-2} + m^2)\hA^a} \equiv
|\triangle^{m^2}_{\hA}|^{-\hf} \ , \label{zmass}
\end{equation}
while for $m^2=0$, $\int \cD \varphi$ corresponds to the gauge
volume and has to be removed from $Z_A$,
\begin{equation}
Z_{massless} = Z_A/\int \cD \varphi =
|\triangle^{m^2=0}_{\hA}|^{-\hf} \, |\triangle_s|^{\hf} \ ,
\label{zless}
\end{equation}
As shown in the above, the anomaly contribution for the massive
vector is written as $-\sqrt{m^2l^2+1} \, v/32\pi^2$, while a
five-dimensional minimally coupled scalar with mass $m^2$ gives
$-\sqrt{m^2 l^2 + 4} \, s/32\pi^2$. Therefore the anomaly
contribution is
\begin{equation}
Z_{massive} \Rightarrow -{\sqrt{m^2l^2+1}\over 32\pi^2} \, v \ ,
\qquad \ Z_{massless} \Rightarrow -{1 \over 32\pi^2} (v - 2 s) \ ,
\end{equation}
identical to the above result obtained in the canonical
formulation.
\par

\section{Graviton}
For the complete proof of (\ref{afsum}), we also have to
investigate the graviton sector since there appears a contribution
from the five-dimensional (ghost) vector field, as shown in Table
3. The Lagrangian for the five-dimensional graviton is obtained by
expanding the Einstein-Hilbert action with cosmological constant
$\Lambda$,
\begin{equation}
{\cal L}_G = \km2 \, \sqrt{g}\, (-R + 2\Lambda) \ .
\end{equation}
w.r.t. $h_{\mu\nu}=g_{\mu\nu}-g^{AdS}_{\mu\nu}$. The term
quadratic in $h_{\mu\nu}$ becomes
\begin{eqnarray}
{\cal L}_{G2} &=& \km2 \sqrt{g}\, [\, \qt\, {\tilde
h}^{\mu\nu}(-g_{\mu\lm}g_{\nu\tau}\, \Box - 2 R_{\mu\lm\nu\tau} +
2 R_{\mu\lm}g_{\nu\tau})\, h^{\lm\tau} - \hf \nabla^\mu {\tilde
h}_{\mu\nu}\nabla^\lm{\tilde h}_\lm^{\ \nu} \nonumber
\\ && \qquad \qquad \qquad - {\tilde h}^{\mu\nu}\, E_{\mu\lm}\, h_\mu^{\ \,
\lm} - \hf \Lambda \, {\tilde h}^{\mu\nu}\, h_{\mu\nu}] \ ,
\end{eqnarray}
where ${\tilde h}_{\mu\nu}= h_{\mu\nu} - \hf g_{\mu\nu}\, h^\lm_{\
\, \lm}$ and $E_{\mu\lm} = R_{\mu\lm} - \qt R g_{\mu\lm}$. We
decompose $h_{\mu\nu}$ into their traceless and trace parts;
$\phi_{\mu\nu} = h_{\mu\nu} - {1 \over 5}g_{\mu\nu}\, h^\lm_{\ \,
\lm}$ and $h = h^\lm_{\ \, \lm}$. With $\phi_{\mu\nu}$ and $h$,
${\cal L}_{G2}$ is expressed in the constant curvature background
as
\begin{eqnarray}
{\cal L}_{G2} &=& {\cal L}_\phi + {\cal L}_h + {\cal L}_{harm.}
\ , \nonumber \\
{\cal L}_\phi &=& \qt \km2 \sqrt{g}\, (\nabla^\lm
\phi^{\mu\nu}\nabla_\lm \phi_{\mu\nu} -2 l^{-2}\,
\phi^{\mu\nu}\phi_{\mu\nu})
\ , \nonumber \\
{\cal L}_h &=& -{3 \over 40} \km2 \sqrt{g}\,
(h\, (-\Box)\, h + 8 l^{-2}\, h^2) \ , \label{Lgh}\\
{\cal L}_{harm.} &=& - \hf \km2 \sqrt{g}\, \nabla^\mu {\tilde
h}_{\mu\nu}\nabla^\lm {\tilde h}_\lm^{\ \nu} = - \hf \km2
\sqrt{g}\,(\nabla^\mu \phi_{\mu\nu} - {3 \over 10} \nabla_\nu
h)(\nabla^\lm \phi_\lm^{\ \nu} -{3 \over 10} \nabla^\nu h) \ .
\nonumber
\end{eqnarray}
We note that the quadratic action $S_{G2} = \int {\cal L}_{G2}$ is
invariant under a {\it finite} transformation $h_{\mu\nu} =
h'_{\mu\nu} + \nabla_\mu V_\nu + \nabla_\nu V_\mu$ or
equivalently,
\begin{equation}
\phi_{\mu\nu} = \phi'_{\mu\nu} + 2\, \nabla_{(\mu}V_{\nu)} \ ,
\qquad h = h' + 2\, \nabla^\mu V_\mu \ ,
\end{equation}
where $\nabla_{(\mu}V_{\nu)}$ is the symmetric and traceless part
of $\nabla_\mu V_\nu$.
\par

Next we add a mass term to ${\cal L}_{G2}$,
\begin{equation}
{\cal L}_{G2} + {\cal L}_{m^2} \equiv {\cal L}_{G2} + \qt \km2
\sqrt{g}\, m^2\, {\tilde h}^{\mu\nu}h_{\mu\nu} = {\cal L}_{G2} +
\qt \km2 \sqrt{g}\, m^2\, (\phi^{\mu\nu}\phi_{\mu\nu} -{3 \over
10}\, h^2) \ ,
\end{equation}
where the form ${\tilde h}^{ab}h_{ab}$ is required (instead of
$h^{\mu\nu}h_{\mu\nu}$) to produce a mass term arising from the
compactification of ten-dimensional Type IIB theory on $S^5$
\cite{Kim}.
\par

The equation of motion for the massive graviton can be cast into
the form,
\begin{equation}
-\Box \phi_{\mu\nu} -2l^{-2}\phi_{\mu\nu} + m^2 \phi_{\mu\nu} = 0
\ , \quad -\Box h + 8l^{-2}h + m^2 h = 0 \ . \label{eqm2}
\end{equation}
Decomposing $\phi_{\mu\nu} = \hph_{\mu\nu} + 3
\nabla_{(\mu}\nabla_{\nu)}h/8(m^2+3l^{-2})$, we can see that
$\phi_{\mu\nu}$ in the first equation can be replaced with
$\hph_{\mu\nu}$ which satisfies the transversal condition
$\nabla^\mu \hph_{\mu\nu}=0$, indicating that the partition
function for the massive graviton is described by the path
integral w.r.t. $\hph_{\mu\nu}$ and $h$.
\par

We decompose the path-integral variable $\phi_{\mu\nu} =
\hph_{\mu\nu} + 2\, \nabla_{(\mu}V_{\nu)}$ such that $\nabla^\mu
\hph_{\mu\nu}=0$ and transform $h$ as $h = \hh + 2\, \nabla^\mu
V_\mu$. Then using the invariance of $S_{G2}[\phi, h] = \int {\cal
L}_{G2}$ under the gauge transformation, we see that the variable
$V_\mu$ only appears in the mass term,
\begin{eqnarray}
Z_G &=& \int \cD\phi \cD h \, e^{-\int ({\cal L}_{G2}+ {\cal
L}_{m^2})} = \int \cD\hph \cD V \cD \hh\, |\triangle_v|^\hf \,
e^{-\int {\cal L}_{G2}[\hph, \hh] - \int {\cal L}_{m^2}[\hph, \hh,
V]} \nonumber \\
&=& \int \cD\hph\, e^{-\int {\cal L}_\hph^{m^2}}\, \int \cD V \cD
\hh\, |\triangle_v|^\hf \, e^{-\int {\cal L}_{\hh, V}^{m^2}} \ ,
\label{zg}
\end{eqnarray}
where
\begin{eqnarray}
{\cal L}_\hph^{m^2} &=& \qt \km2 \sqrt{g}\, [\nabla^\lm
\hph^{\mu\nu}\nabla_\lm \hph_{\mu\nu} +(m^2 -2 l^{-2})\,
\hph^{\mu\nu}\hph_{\mu\nu}]
\ , \label{Lhph}\\
{\cal L}_{\hh, V}^{m^2} &=& -{3 \over 25}\, \km2 \sqrt{g}\, [\hh\,
(-\Box)\, \hh + 5 l^{-2}\, \hh^2 ] \nonumber
\\ &+& \qt \km2 \sqrt{g}\, m^2\, [2\, V_\mu \triangle_v^{\mu\nu} V_\nu
-{3 \over 10}\, (\hh + 2\nabla^\mu V_\mu)^2] \ , \label{Lhv}
\end{eqnarray}
and the operator $\triangle_v$ acting on a vector field $V_\nu$ is
\begin{equation}
\triangle_{v\,\mu}^{\ \ \ \nu} V_\nu \equiv - 2\, \nabla^\nu
\nabla_{(\mu}V_{\nu)} = (-\Box_\mu^{\ \nu} + 4l^{-2}\dl_\mu^{\
\nu} - {3 \over 5} \nabla_\mu \nabla^\nu)\, V_\nu \ . \label{triv}
\end{equation}
which can be factorized under the decomposition of the path
integral variable $A_\mu = \hA_\mu + \pd_\mu \varphi$ with
$\nabla^\mu \hA_\mu = 0$ as
\begin{eqnarray}
|\triangle_v|^{-\hf} &=& \int \cD A \, e^{-\hf \int \sqrt{g}\,
A_\mu \triangle_v^{\mu\nu} A_\nu} \nonumber \\ &=& \int \cD \hA
e^{-\hf \int \sqrt{g}\, \hA^\mu (-\Box + 4l^{-2})\hA_\mu}\, \int
\cD \varphi\, |\triangle_s|^\hf \, e^{-{4 \over 5} \int \sqrt{g}\,
\varphi (\triangle_s^2 +
5l^{-2}\triangle_s)\varphi} \label{jacob} \\
&=& |\triangle^{m^2=8l^{-2}}_{\hA}|^{-\hf}\, |\triangle_s +
5l^{-2}|^{-\hf} \ , \nonumber
\end{eqnarray}
where $|\triangle^{m^2}_{\hA}|^{-\hf}$ is given in (\ref{zmass}).
\par

For $m^2 \not= 0$, the path-integral for $V_\mu$ and $\hh$ in
(\ref{zg}) is performed in a similar way after decomposing $V_\mu
= \hV_\mu + \pd_\mu \varphi$ s.t. $\nabla^\mu \hV_\mu = 0$,
\begin{eqnarray}
&&\int \cD V \cD \hh\, |\triangle_v|^\hf \, e^{-\int {\cal
L}_{\hh, V}^{m^2}} \nonumber \\ &&=
|\triangle^{m^2=8l^{-2}}_{\hA}|^\hf\, |\triangle_s + 5l^{-2}|^\hf
\, \int \cD \hV e^{-\hf \km2 m^2\,
\int \sqrt{g}\, V_\mu (-\Box + 4l^{-2}) V^\mu}\, \times \nonumber \\
&& \times \int \cD \varphi \cD \hh \, |\triangle_s|^\hf\,
e^{-{\beta \over 4}\km2 \int \sqrt{g}\, [{8 \over 5}\hh
\triangle_s \hh + (8l^{-2}+m^2)\hh^2 -{20 \over 3}m^2\, \varphi
(\triangle_s^2 + 8l^{-2}\triangle_s)\varphi -4m^2\, \hh
\triangle_s
\varphi]} \nonumber \\
&&= |\triangle_s + 5l^{-2}|^\hf\, |\triangle_s|^\hf\,
\left|\begin{array}{cc} {8 \over 5}\triangle_s + 8l^{-2}+m^2 &
-2m^2\, \triangle_s \\ -2m^2\, \triangle_s & -{20 \over 3}m^2\,
(\triangle_s^2 + 8l^{-2}\triangle_s) \end{array} \right|^{-\hf}
\nonumber
\\ && \sim |\triangle_s + 8l^{-2} +m^2|^{-\hf} \ , \label{pivh}
\end{eqnarray}
while the $\hph$-integral in (\ref{zg}) is denoted as
\begin{equation}
\int \cD\hph\, e^{-\int {\cal L}_\hph^{m^2}} = \int \cD\hph\,
e^{-\qt \km2 \int \sqrt{g}\, \hph^{\mu\nu}(-\Box -2 l^{-2} +
m^2)\hph_{\mu\nu}} \equiv |\triangle_\hph^{m^2}|^{-\hf} \ ,
\end{equation}
which gives
\begin{equation}
Z_G = |\triangle_\hph^{m^2}|^{-\hf}\, |\triangle_s + 8l^{-2}
+m^2|^{-\hf} \ .
\end{equation}
Compared with (\ref{eqm2}), we see that the determinant
$|\triangle_s + 8l^{-2} +m^2|^{-\hf}$ stems from the trace part
$h$, which however is coupled to other scalars $\pi$ and $b$ in
ten dimensional Type IIB theory on $S^5$ and indeed is given by
$\pi$ as $h = (16/15)\pi$ for the massive case ($m^2 =
k(k+4)l^{-2}$, $k \geq 1$) \cite{Kim}. As the mass spectrum of
$\pi$ and $b$ is listed in Table III of \cite{Kim} and has already
been counted in our sum of KK-modes, the anomaly contribution from
the massive graviton corresponds to the $\hph$-mode,
$|\triangle_\hph^{m^2}|^{-\hf}$.
\par

For $m^2=0$, we have to take into account the trace mode. From
(\ref{zg}), (\ref{Lhph}), (\ref{Lhv}) with $m^2=0$, we have
\begin{eqnarray}
Z_G &=& \int \cD\hph\, e^{-\int {\cal L}_\hph^{m^2}}\, \int \cD
\hh\, e^{-{2\beta \over 5} \km2 \int \sqrt{g}\, \hh( -\Box + 5
l^{-2})\hh}\, \int \cD V\, |\triangle_v|^\hf \nonumber \\
&=& |\triangle_\hph^{m^2=0}|^{-\hf}\, |\triangle_s +
5l^{-2}|^{-\hf}\, |\triangle^{m^2=8l^{-2}}_{\hA}|^\hf\,
|\triangle_s + 5l^{-2}|^\hf\, \int \cD V \nonumber \\
&=& |\triangle_\hph^{m^2=0}|^{-\hf}\,
|\triangle^{m^2=8l^{-2}}_{\hA}|^\hf\, \int \cD V \ ,
\end{eqnarray}
where $\int \cD V$ corresponds to the gauge volume of the massless
graviton theory and thus has to be discarded from $Z_G$.
\par

Solving the Schr\"{o}dinger equation, we see that the anomaly
contribution from the traceless and transversal $\hph$-mode,
$|\triangle_\hph^{m^2}|^{-\hf}$ is $-\sqrt{m^2l^2+4} \, g/32\pi^2$
with a mass-independent parameter $g$, while as noted in Section
2, $|\triangle^{m^2=8l^{-2}}_{\hA}|^{-\hf}$ gives $-\sqrt{8+1}
v/32\pi^2 = -3v/32\pi^2$. Therefore the anomaly contributions from
the massive and massless graviton are
\begin{equation}
Z_G^{massive} \Rightarrow - {\sqrt{m^2l^2+4} \over 32\pi^2}\, g \
, \qquad Z_G^{massless} \Rightarrow -{1 \over 32\pi^2}\, (2g -3v)
\ , \label{zsum}
\end{equation}
where $v = v_0+2s-2s_0$, which completes the proof of
(\ref{afsum}).
\par

\section{Anti-symmetric Tensor}
Finally, we consider the theory of a massive anti-symmetric field
$B_{\mu\nu}$. The five-dimensional Lagrangian is given as
\begin{eqnarray}
{\cal L}_B &=& \sqrt{g}\, [\, {1 \over 24}\,
F_{\mu\nu\lm}F^{\mu\nu\lm} + \hf \, m^2 B_{\mu\nu} B^{\mu\nu} \,]
\\ &=& \hf \sqrt{g}\, [(\nabla_\lm B_{\mu\nu})^2 -2(\nabla^\mu
B_{\mu\nu})^2 + B^{\lm\tau}(2R_\lm^{\ \, \mu}\dl_\tau^{\ \, \nu} -
R_{\lm\tau}^{\ \ \, \mu\nu})B_{\mu\nu} + m^2 B^{\mu\nu}B_{\mu\nu}]
\ . \nonumber
\end{eqnarray}
As in the vector case, we decompose $B_{\mu\nu}$ into the
transversal part $\hB_{\mu\nu}$ with $\nabla^\mu \hB_{\mu\nu}=0$
and the gauge mode $\pd_{[\mu}\zeta_{\nu]}$. Then the partition
function is written as
\begin{equation}
Z_B = \int {\cal D}B \, e^{-\int {\cal L}_B} = \int \cD \hB \,
e^{-\int \sqrt{g}[{1 \over 24}F^2(\hB) + \hf m^2 \hB^2]}\, \int
\cD \zeta |\triangle_\zeta|^{\hf} \, e^{-\qt m^2 \int \sqrt{g}\,
\zeta_\mu \triangle_\zeta^{\mu\nu} \zeta_\nu} \ , \label{zb}
\end{equation}
where the determinant of the operator $\triangle_\zeta^{\mu\nu}=
-\Box^{\mu\nu} -4l^{-2}g^{\mu\nu} + \nabla^\mu \nabla^\nu$ is
expressed under the decomposition $\zeta_\mu =\hzt_\mu + \pd_\mu
\phi$ with $\nabla^\mu \hzt_\mu = 0$,
\begin{eqnarray}
|\triangle_\zeta|^{-\hf} &=& \int \cD \zeta \, e^{-\hf \int
\sqrt{g}\, \zeta_\mu \triangle_\zeta^{\mu\nu} \zeta_\nu} = \int
\cD \hzt \, e^{-\hf \int \sqrt{g}\, \hzt_\mu
\triangle_\zeta^{\mu\nu} \hzt_\nu} \, \int \cD \phi \,
|\triangle_s|^{\hf} \nonumber \\
&=& |\triangle^{m^2=0}_{\hA}|^{-\hf} \, |\triangle_s|^{\hf}\, \int
\cD \phi \ ,
\end{eqnarray}
where $|\triangle^{m^2=0}_{\hA}|^{-\hf}$ is given in
(\ref{zmass}), showing that the determinant diverges due to the
gauge invariance of $\triangle_\zeta^{\mu\nu}$ under $\dl
\zeta_\mu = \pd_\mu \, \phi$. For $m^2\not= 0$, however the
determinant is suppressed by the path integral w.r.t. $\zeta$ and
the remaining part in (\ref{zb}) contributes to the Weyl anomaly
as
\begin{equation}
Z_B^{massive} = \int \cD \hB \, e^{-\int \sqrt{g}[{1 \over
24}F^2(\hB) + \hf m^2 \hB^2]} \ \Rightarrow \ -{|m| \over
32\pi^2}\, b
 \ , \label{mb}
\end{equation}
where $b$ is a mass independent parameter. For $m^2=0$, the path
integral for $\hB_{\mu\nu}$ gives no effect to the anomaly,
although there remains the Jacobian in (\ref{zb}) giving
\begin{equation}
\int \cD \zeta\, |\triangle_\zeta|^{\hf} =
|\triangle^{m^2=0}_{\hA}|^{\hf} \, |\triangle_s|^{-\hf}\, (\int
\cD \phi)^{-1}\, \int \cD \hzt\, \cD \phi \, |\triangle_s|^{\hf} =
|\triangle^{m^2=0}_{\hA}|^{\hf}\, \int \cD \hzt \ ,
\end{equation}
where $\int \cD \hzt$ for the constrained variable $\hzt_\mu$ with
$\nabla^\mu \hzt_\mu = 0$ corresponds to the gauge volume of the
theory and has to be removed, which leads to the contribution to
the anomaly
\begin{equation}
Z_B^{massless} = |\triangle^{m^2=0}_{\hA}|^{\hf} \ \Rightarrow \
-{1 \over 32\pi^2}(-v)  \ .
\end{equation}
Thus the vector parameter arising from the massless $B_{\mu\nu}$
is also given by $v=v_0-2s_0+2s$. Note that, however, the massless
anti-symmetric tensor only appears in the doubleton
supermultiplet, which consists of the first three fields with
$p=1$ in Table 1, that is, $A^{(1)}_{\mu\nu}$ with
$\triangle-2=0$. As the doubleton is known to correspond to the
center-of-mass degree of freedom of the boundary theory, that is,
the $U(1)$ factor of $U(N) = SU(N) \times U(1)$ Super-Yang-Mills,
we did not count the doubleton sector in the summation
(\ref{afsum}).

\section{Diagonalisation of the Spectrum}
To construct the spectrum we reduce the ten-dimensional Type IIB
Supergravity action about the $AdS_5\times S^5$ background,
expanding in $S^5$ spherical harmonics to obtain a
five-dimensional action on $AdS_5$. For the purposes of
calculating the one-loop Weyl anomaly, we need the quadratic part
of the 5d action. At the time that we began this work, this had
not been calculated \footnote{A construction of the quadratic
action has since been given in \cite{frolov}. This is equivalent
to the action we constructed and describe here, although the
details of the derivation are different.} , so we constructed an
action that reproduces the field equations of ten-dimensional Type
IIB Supergravity, expanded in $S^5$ harmonics. The difficulties
associated with a Lagrangian description of self-dual field
strengths can be avoided by expanding in $S^5$ harmonics before
constructing the Lagrangian: thus the action we construct is local
in $AdS_5$ but not in the ten-dimensional space.

\subsection{Graviton}

To begin with we exclude couplings between the metric and gauge
field, and consider the five-dimensional equations of motion that
arise from pure ten-dimensional gravity on the $AdS_5\times S^5$
background. Decomposing the ten-dimensional metric into backround
values and fluctuations as $g_{mn}=\dot g_{mn}+h_{mn}$, we have
\be \dot R_{\mu\nu} = {4\over l^2}g_{\mu\nu},\qquad \dot
R_{\alpha\beta} = - {4\over l^2}g_{\alpha\beta},
\label{einstein}\ee where $g_{\mu\nu}$ is the deformed $AdS_5$
metric (\ref{ads1}), and $g_{\alpha\beta}$ is the metric of $S^5$.
Indices will be raised and lowered with these metrics.

The ten-dimensional Einstein term can be written to quadratic
order as \bq \sqrt{g} R|_{\dot g+h}&=&({\sqrt{\dot g}}\dot
R)-h^{mn}({\sqrt{\dot g}}\dot R_{mn}
-\hf{\sqrt{\dot g}}\dot g_{mn}\dot R)-\nonumber\\
&&-\hf h^{mn} \Big( \sqrt{\dot g} E_{mn}^{\ \ pq} h_{pq} + \hf
\dot g^{pq} h_{pq} ({\sqrt{\dot g}}\dot R_{mn}-\hf{\sqrt{\dot
g}}\dot g_{mn}\dot R)
-\nonumber\\
&&-\hf\sqrt{\dot g} h_{mn} \dot R  +  \hf\sqrt{\dot g} \dot g_{mn}
 h ^{pq} \dot R_{pq}-\hf\sqrt{\dot g} \dot g_{mn} \dot g^{pq} E_{pq}^{\ \ rs}
 h_{rs}
\Big), \label{quad}\eq where Roman indices refer to
ten-dimensional coordinates, and $E_{mnpq}$ is a second-order
differential operator. Call the part quadratic in $h_{mn}$ $S_2$.
We will impose the gauge conditions \be
D^{\alpha}h_{\alpha\beta}=0, \qquad D^{\alpha}h_{\alpha\mu}=0
\label{gc}.\ee {}From the ten-dimensional Einstein equations
(\ref{einstein}) we have \bq \dot R &=&0\n \dot R^{mn} h_{mn} &=&
{4\over l^2}(h_{\mu}^{\mu} - h_{\alpha}^{\alpha}). \eq Also, \bq -
h^{mn} \dot R_{mrsn} h^{rs} &=& {1\over l^2}(h^{\mu \nu} h_{\mu
\nu} - h_{\mu}^{\mu} h_{\upsilon}^{\upsilon})  + {1\over
l^2}(h^{\alpha \beta} h_{\alpha \beta} - h_{\alpha}^{\alpha}
h_{\beta}^{\beta})\n h^{mn} \dot R_m^r h_{nr} &=& {4\over
l^2}(h^{\mu \upsilon} h_{\mu \upsilon} - h^{\alpha \beta}
h_{\alpha \beta})\n
 \dot g^{pq} E_{pq}^{\ \ rs} h_{rs} &=& \hf(2\Box_x +
\Box_y) h_m^m - D_m D_n h^{mn}\n h^{mn} E_{mn}^{\ \
pq}h_{pq}&=&h^{mn}(\hf(\Box_x+\Box_y)+D_mD_nh^r_r-D_mD^rh_{rn}),
\eq so that \bq \frac{-2S_2}{\sqrt{\dot g}} &=& \hf h^{mn} (\Box_x
+ \Box_y) h_{mn} - \hf h_m^m(\Box_x + \hf \Box_y) h_n^n +\n &&+
h^{mn} D_m D_n h_r^r - h^{mn} D_m D^r h_{rn} \n &&+ {5\over l^2}
h^{\mu \upsilon} h_{\mu \upsilon} - {5\over l^2} h^{\alpha \beta}
h_{\alpha \beta} + {3\over l^2} h_{\mu}^{\mu}
h_{\upsilon}^{\upsilon} - {3\over l^2} h_{\alpha}^{\alpha}
h_{\beta}^{\beta}. \eq Taking a variation of the action
(\ref{quad}) with respect to $h^{mn}$ gives the equation of motion
(neglecting mass terms) \bq 2\dot R_{mn} - \dot g_{mn} \dot R &=&
(\Box_x + \Box_y) h_{mn}- \dot g_{mn} (\Box_x + \hf \Box_y) h_p^p
+ D_{(m} D_{n)} h_p^p \n &&+ \dot g_{mn} D^p D^q h_{pq} - 2D_{(m}
D^r h_{n)r}.\eq This contracts to \be \dot R =  (\Box_x + \hf
\Box_y) h_p^p - D^p D^q h_{pq}, \ee whence \bq  2 \dot R_{(\mu
\nu)} &=& (\Box_x + \Box_y) h'_{\mu \nu} + D_{(\mu} D_{\nu)}
h'^\mu_\mu - 2 D_{(\mu} D^{\rho} h'_{\nu)\rho} \n &=&
2E^{1.1}_{(\mu \nu)}\label{e1.1}. \eq Here $h'_{\mu\nu}$ is
defined by a linearised Weyl shift:
$h_{\mu\nu}=h'_{\mu\nu}-{1\over3}g_{\mu\nu}h^\alpha_\alpha$. Round
brackets indicate that an index pair is symmetrised with the trace
removed. In writing (\ref{e1.1}) we made use of the gauge
conditions (\ref{gc}). Also, we have \bq 2 g^{\mu \nu} R_{\mu \nu}
&=& (2\Box_x + \Box_y) h^\alpha_\alpha - \frac{5}{3} (\Box_x +
\Box_y) h'^\mu_\mu - 2 D^{\mu} D^{\nu} h'_{\mu \nu} \n
&=&10E^{1.2},\eq

\bq 2R_{(\alpha \beta)} &=& (\Box_x + \Box_y) h_{(\alpha \beta)} +
(D_{(\alpha} D_{\beta)} h^\alpha_\alpha - \frac{16}{15}
D_{(\alpha} D_{\beta)} h'^\mu_\mu)-2D_{(\alpha}D^\mu
h_{\beta)\mu}\n &=& 2E^{3.1}_{\alpha \beta} + 2E^{3.2}_{\alpha
\beta} - 2E^{3.3}_{\alpha \beta}, \eq

\bq 2g^{\alpha \beta} R_{\alpha \beta} &=& (\Box_x - \frac{1}{15}
\Box_y) h'^\mu_\mu + \Box_y h^\alpha_\alpha \n &=& 10 E^{3.4}, \eq

\bq 2R_{\mu\alpha}&=&\left(\delta^\nu_\mu(\Box_x+\Box_y)-D_\mu
D^\nu\right)h_{\nu\alpha}+D_\alpha\left(D_\mu(h^\alpha_\alpha
+{8\over15}h'^\mu_\mu)-D^\nu h'_{\mu\nu}\right) \n &=&
2E^{2.1}_{\mu\alpha}+2E^{2.2}_{\alpha\mu}\label{e2}.\eq The
equations of motion arising from the action $S_2$ imply the
vanishing of the quantities (\ref{e1.1})-(\ref{e2}). We expand
everything in $S^5$ spherical harmonics as follows: \bq
h'_{\mu\nu}&=&\sum H_{\mu\nu}(x)Y(y),\qquad h_{\mu\alpha}=\sum
B_\mu(x) Y_\alpha(y), \n h_{(\alpha\beta)}&=&\sum\phi(x)
Y_{(\alpha\beta)}(y),\qquad h^\alpha_\alpha=\sum\pi(x)Y(y). \eq
When we include the couplings to the antisymmetric field, the
equations of motion (\ref{e1.1})-(\ref{e2}) give the equations of
motion E1.1-E3.4 in Table II of \cite{Kim}, each being
proportional to a single spherical harmonic. We will refer
throughout to the equations in Table II of (\cite{Kim}) since they
give a convenient way of checking the coefficients of coupling and
mass terms.

\subsection{Antisymmetric Tensor}

Again we begin by excluding couplings to the metric, and seek to
construct an action that reproduces the equations of motion in
Table II of \cite{Kim}. We decompose the ten-dimensional
four-index antisymmetric tensor in terms of background values and
fluctuations as $A_{mnpq}=\dot A_{mnpq}+a_{mnpq}$ and expand the
fluctuations $a_{mnpq}$ in spherical harmonics as follows:

\bq a_{\mu\nu\rho\sigma}&=&\sum b_{\mu\nu\rho\sigma}(x)Y(y), \n
a_{\mu\nu\rho\alpha}&=&\sum b_{\mu\nu\rho}(x)Y_{\alpha}(y), \n
a_{\mu\nu\alpha\beta}&=&\sum b_{\mu\nu}(x)Y_{[\alpha\beta]}(y), \n
a_{\mu\alpha\beta\gamma}&=&\sum
\phi_{\mu}(x)\epsilon_{\alpha\beta\gamma}^{\delta\epsilon}D_\delta
Y_{\epsilon}(y), \n a_{\alpha\beta\gamma\delta}&=&\sum
b(x)\epsilon_{\alpha\beta\gamma\delta}^{\epsilon}D_\epsilon Y(y).
\eq Consider the action \be S^{40} = b \epsilon^{\mu \nu \rho
\sigma \tau} \partial_{\mu} b_{\nu \rho \sigma \tau} + 12 b \Box_y
b -\hf b^{\mu \nu \rho \sigma} b_{\mu \nu \rho \sigma}. \ee
Varying $b$ gives M1: \be 5 \partial_{[\mu} b_{\nu \rho \sigma
\tau]} - \epsilon_{\mu \nu \rho \sigma \tau} \Box_y b. \ee Varying
$b^{\mu \nu \rho \sigma}$ gives M2.2: \be -\epsilon^{\tau \mu \nu
\rho \sigma}\partial_{\tau}b - b_{\mu \nu \rho \sigma}. \ee Now
consider the action \be S^{31} = \hf b^{\mu \nu \rho} b_{\mu \nu
\rho} + b^{\mu \nu \rho} \epsilon_{\mu \nu \rho}^{\sigma \tau}
\partial_{\sigma} \phi_{\tau} + 3\phi^{\tau} \triangle_y
\phi_{\tau}. \ee Varying $b^{\mu \nu \rho}$ gives M3.2: \be b_{\mu
\nu \rho} + \epsilon_{\mu \nu \rho}^{\sigma \tau}
\partial_{\sigma} \phi_{\tau},\ee
while varying $\phi^\tau$ gives \be - \epsilon_{\mu \nu
\rho}^{\sigma \tau} \partial_{\sigma} b^{\mu \nu \rho} + 6
\triangle_y \phi_{\tau}, \ee which is equivalent to M2.1: \be 4
\partial_{[\mu} b_{\nu \rho \sigma]} + \epsilon_{\mu \nu \rho
\sigma}^{\tau} \triangle_y \phi_{\tau}. \ee Finally, consider the
action \be S^{22}=b^+_{\mu\nu}\partial_\rho
b^-_{\sigma\tau}\epsilon^{\mu\nu\rho\sigma\tau}-ib^{+\mu\nu}
\sqrt{-\Delta_y}b^-_{\mu\nu}. \ee Varying $b^+_{\mu\nu}$ gives \be
\partial_\rho
b^-_{\sigma\tau}\epsilon^{\mu\nu\rho\sigma\tau}-i\sqrt{-\Delta_y}
b^-_{\mu\nu},\ee which corresponds to M3.1.

\subsection{Gravitational Couplings}

To generate the correct couplings to gravity in the equations of
motion for the antisymmetric tensor we make the modifications

\be S^{40}\rightarrow S^{40}_{int}=S^{40}+{12\over l}Hb-{32\over
l} \pi b, \ee \be S^{31}\rightarrow S^{31}_{int}=S^{31}+{6\over l}
\phi^\tau B_\tau, \ee where $H=H^\mu_\mu$. The action at this
stage is \be S_2+A_1S^{40}_{int}+A_2S^{31}_{int}+A_3S^{22}_{int},
\ee and the normalisations can be fixed by considering the terms
in the Einstein equations generated by the interactions. The
contribution to $E^{1.2}$ is $-{12\over l}A_1b$. Now in equation
E1.2 of \cite{Kim} we find a term

\be {1\over3l}\epsilon^{\mu \nu \rho \sigma \tau} \partial_{\mu}
b_{\nu \rho \sigma \tau}= -{8\over
l}\left({1\over2l}H-{4\over3l}\pi+\Box_y b\right), \ee where we
used the equation of motion M1. So if ${12\over l}A_1={8\over
l}\Box_y$, the correct coupling is generated, along with some mass
terms. Note that $\Box_y$ has zero modes. This choice also
generates the correct coupling in equation M3.4.

The contribution of the interaction terms to $E^{2.1}_{\mu\alpha}$
is ${3\over l}A_2\phi_\mu$. In equation E2.1 we find the
interaction terms \be -\left(-{2\over3l}\epsilon_{\mu}^{\nu \rho
\sigma \tau}\partial_{\nu} b_{\rho \sigma \tau}+{4\over
l}\Delta_y\phi_\mu\right), \ee which we can rewrite with the help
of M2.1 as

\be -{8\over l}\Delta_y\phi_\mu+{4\over l^2}B_\mu. \ee So we can
take $A_2=-{4\over3}\Delta_y$. The normalisation of $A_3$ does not
need to be fixed, as the action is diagonal in the field
$b^\pm_{\mu\nu}$.

\subsection{Mass Terms}

To calculate the mass terms we will drop the existing mass terms
from $S_2$ and add a mass term to the action:

\bq S_{mass}&=&-\hf B_1h^{\mu\nu}h_{\mu\nu}-\hf
B_2(h^{\mu\nu}g_{\mu\nu})^2-\hf B_3
h^{\alpha\beta}h_{\alpha\beta}-\n &&-\hf
B_4(h^{\alpha\beta}g_{\alpha\beta})^2-Ch^\alpha_\alpha
h^\mu_\mu-\hf D h^\alpha_\mu h^\mu_\alpha. \eq The coefficients of
the mass terms are easily determined from the Einstein equations.
From E3.1 we find $B_3=-{1\over l^2}$, while E1.1 gives
$B_1={1\over l^2}$. E3.4, after a bit of calculation, leads to the
equations

\be 5B_2-3C+{1\over l^2}=0, \qquad 3B_4-5C+{25\over l^2}=0, \ee
while E1.2, rewritten as before with the help of M1 to eliminate
the four-index antisymmetric field, gives

\be 3B_2-5C+{7\over l^2}=0,\qquad 5B_4-3C+{31\over l^2}=0. \ee
These have the consistent solution $B_2={1\over l^2}$, $C={2\over
l^2}$, $B_4=-{5\over l^2}$. Finally, E2.1 gives $D=-{6\over l^2}$,
where this includes an extra $-2$ that arises because we wrote the
action $S_2$ in terms of $\Box_y$ instead of $\Delta_y$.

The final form of $S_{mass}$, is then

\be S_{mass}={1\over l^2}=\left(-\hf H^{(\mu\nu)}H_{(\mu\nu)}+\hf
\phi^2+6\phi^\mu \phi_\mu-{3\over 5}H^2+{64\over15}\pi^2\right).
\ee where we have written everything in terms of harmonically
expanded fields, but the explicit dependence on spherical
harmonics has been suppressed.

\subsection{Diagonalisation}

Writing the complete action so far in terms of the expanded fields
gives:

\bq S&=&-\qr H^{(\mu\nu)}(\Box_x+\Box_y)H_{(\mu\nu)}-\hf D_\mu
H^{(\mu\nu)}D^\rho
H_{(\rho\nu)}-{1\over2l^2}H^{(\mu\nu)}H_{(\mu\nu)}-\qr
\phi(\Box_x+\Box_y)\phi+\n && +{1\over2l^2}\phi^2-\hf
\phi^\mu(\Box_x+\Box_y)\phi_\mu-\hf D_\mu\phi^\mu
D_\nu\phi^\nu+{6\over
l^2}\phi^\mu\phi_\mu+H\left({3\over25}\Box_x+{1\over5}\Box_y\right)H-\n
&&
-{3\over5l^2}H^2+\pi\left({2\over225}\Box_x-{2\over15}\Box_y\right)\pi+{64\over15l^2}\pi^2
-{8\over30}H\Box_y\pi-{3\over10}HD^\mu D^\nu H_{(\mu\nu)}+\n &&
+{2\over3}\tilde
b\left(\epsilon^{\mu\nu\rho\sigma\tau}\partial_\mu
b_{\nu\rho\sigma\tau}-12(1-\delta_{I,0})\tilde b+{12\over
l}H-{32\over
l}\pi\right)-{1\over3}b^{\mu\nu\rho\sigma}\Box_yb_{\mu\nu\rho\sigma}-\n
&&-{8\over3}\left(\hf
b^{\mu\nu\rho}\Delta_yb_{\mu\nu\rho}+b^{\mu\nu\rho}\epsilon_{\mu\nu\rho}^{\
\ \
\sigma\tau}\partial_\sigma\Delta_y\phi_\tau-3\Delta_y\phi^\tau\Delta_y\phi_\tau+{6\over
l}\Delta_y\phi^\tau B_\tau\right)+\n &&
+A_3\left(b^+_{\mu\nu}\partial_\rho
b^-_{\sigma\tau}\epsilon^{\mu\nu\rho\sigma\tau}-ib^{+\mu\nu}\sqrt{-\Delta_y}
b^-_{\sigma\tau}\right).\label{action} \eq We have defined $\tilde
b=\Box_y b$ for future convenience, and the delta function
$\delta_{I,0}$ is equal to 1 for the spherical harmonic for which
$\Box_y$ has eigenvalue 0. In \cite{Kim} the fields
$b_{\mu\nu\rho}$ and $b_{\mu\nu\rho\sigma}$ were algebraically
eliminated from the equations of motion. This is equivalent to
shifting the fields:
$b_{\mu\nu\rho}=b^q_{\mu\nu\rho}+b^c_{\mu\nu\rho}$ and
$b_{\mu\nu\rho\sigma}=b^q_{\mu\nu\rho\sigma}+b^c_{\mu\nu\rho\sigma}$,
where the ``classical'' parts satisfy equations of motion
corresponding to M2.1 and M2.2:

\be \Delta_yb^c_{\mu\nu\rho}+\epsilon_{\mu\nu\rho}^{\ \ \
\sigma\tau}\partial_\sigma\Delta_y\phi_\tau=0, \ee \be
b^c_{\mu\nu\rho\sigma}+\epsilon_{\mu\nu\rho\sigma}^{\ \ \ \
\tau}\partial_\tau\tilde b=0. \ee The quantum parts decouple and
are non-dynamical, while the part of the action involving
$\phi^\tau$ and $\tilde b$ becomes

\bq
&&8\left(-2\partial_{[\sigma}\phi_{\tau]}\Delta_y\partial^{[\sigma}\phi^{\tau]}+\Delta_y\phi^\tau\Delta_y\phi_\tau
-{2\over l}\Delta_y\phi^\tau B_\tau\right)-\n && -8\tilde
b\left((-\Box_x\Box^{-1}_y-1)\tilde b-{1\over
l}H+{8\over3l}\pi\right). \eq In this expression we have assumed
for the moment that the eigenvalue of $\Box_y$ is non-zero. Acting
on $\phi_\mu$ and $B_\mu$, $\Delta_y$ has the eigenvalues
$-{1\over l^2}(k+1)(k+3)$, where $k=1,2,3,\ldots$. Also,
$\Box_y=\Delta_y+{4\over l^2}$. We can diagonalise the $\phi_\mu$,
$B_\mu$ system by putting

\be A^{(1)}_\mu=B_\mu-{4\over l}(k+3)\phi_\mu,\qquad
A^{(2)}_\mu=B_\mu-{4\over l}(k+1)\phi_\mu,\ee and the masses take
the expected values $M^2l^2=(k^2-1)$ and $M^2l^2=(k+3)(k+5)$
respectively.

To diagonalise the graviton, we make the orthogonal decomposition

\be H_{(\mu\nu)}=\hat
h_{\mu\nu}+D_{(\mu}\Lambda_{\nu)}+D_{(\mu}D_{\nu)}\Box_x^{-1}\tilde\phi,\label{dec}
\ee where the components satisfy the transversality conditions
$D^\mu\hat h_{\mu\nu}=D^\mu\Lambda_\mu=0$. Inserting this
decomposition into (\ref{action}) there are no cross-terms. The
$\tilde\phi$ part of the action becomes

\be
\tilde\phi\left({3\over25}\Box_x+\left(-{3\over5l^2}-{1\over5}\Box_y\right)+{1\over
l^2}
\Box^{-1}_x\Box_y\right)\tilde\phi-{6\over5}H\left({1\over5}\Box_x-{1\over
l^2}\right)\tilde\phi. \label{pa}\ee To get rid of the
$\Box_x^{-1}$ dependence we introduce an additional field $\psi$.
(\ref{pa}) becomes

\be
\tilde\phi\left({3\over25}\Box_x-{3\over5l^2}-{1\over5}\Box_y\right)\tilde\phi+2\tilde\phi\psi-l^2\psi\Box_x\Box^{-1}_y\psi
-{6\over5}H\left({1\over5}\Box_x-{1\over
l^2}\right)\tilde\phi.\label{pa2} \ee The rest of the scalar part
of the action is

\bq &&
H\left({3\over25}\Box_x-{3\over5l^2}+{1\over5}\Box_y\right)H+\pi\left({2\over225}\Box_y
-{2\over15}\Box_x+{64\over15l^2}\right)\pi-\n &&
-{8\over30}H\Box_y\pi-8\tilde b\left((-\Box_y^{-1}\Box_x-1)\tilde
b-{1\over l}H+{8\over3l}\pi\right).\label{scal}\eq Making the
change of variables $H=\phi_1+\phi_2$, $\tilde\phi=\phi_1-\phi_2$,
the total scalar action (\ref{scal})+(\ref{pa2}) can be written

\bq S_{scalar}&=&\phi_1\left({4\over5}\Box_y\phi_2+2\psi+8\tilde
b-{4\over15}\Box_y\pi\right)
+\phi_2\left({12\over25}\Box_x-{12\over5l^2}\right)\phi_2-\n &&
-{1\over
l^2}\psi\Box_x\Box_y^{-1}\psi+\pi\left({2\over225}\Box_y-{2\over15}\Box_x+{64\over15l^2}\right)\pi+8\tilde
b(\Box_x\Box_y^{-1})\tilde b+ \n &&
+\phi_2\left(-{4\over15}\Box_y\pi+{8\over l}\tilde
b-2\psi\right)-{64\over3l}\tilde b\pi. \eq The integral over
$\phi_1$ now imposes the condition

\be {4\over5}\Box_y\phi_2+2\psi+8\tilde
b-{4\over15}\Box_y\pi=0,\ee which can be used to eliminate the
field $\psi$.

Changing variables from $(\phi_2,\pi,\tilde b)$ to

\be
(X,Y,Z)=\left({5\over2}\left(\phi_2-{1\over3}\pi\right)/\sqrt{5+l^2\Box_y},
i{5\over\sqrt2}\left({5\over3}\pi-2\phi_2\right),i\sqrt{\Box_y\over8}\left(\tilde
b+{l^2\over5}\Box_y\phi_2\right)\right), \ee we find that the
kinetic term is diagonal in $(X,Y,Z)$. Diagonalising the mass
matrix then gives mass eigenvalues $M^2l^2=k(k-4)$, $(k+4)(k+8)$,
and $5$. All this assumed that the eigenvalue of $\Box_y$ was
non-zero, so $k=2,3,\ldots$. The modes on which $\Box_y=0$ will be
considered shortly.

The part of the action involving $\Lambda_\mu$ can be written as

\be {1\over8}\Box_y\Lambda\left(\Box_x-{4\over l^2}\right)\Lambda,
\ee and on spherical harmonics for which $\Box_y\ne 0$ the
integration over $\Lambda$ just cancels the Jacobian for the
change of variables (\ref{dec}). For the mode with $\Box_y=0$ the
Jacobian is not cancelled. In either case, if we put together the
actions for $h_{(\mu\nu)}$ $\Lambda_\mu$ and the scalar of mass
$5$, we get the correct quadratic action for a five-dimensional
massive (or massless) graviton, as considered in Section 4.

Finally we consider the modes for which the eigenvalue of $\Box_y$
vanishes. In this case, after the decomposition (\ref{dec}), the
scalar part of the action becomes

\be {3\over25}(H-\Box_x\tilde\phi)\left(\Box_x-{5\over
l^2}\right)(H-\Box_x\tilde\phi)-{2\over15}\pi\left(\Box_x-{32\over
l^2}\right)\pi+{2\over3}\tilde
b\epsilon^{\mu\nu\rho\sigma\tau}\partial_\mu\hat
b_{\nu\rho\sigma\tau}, \ee where $\hat b_{\nu\rho\sigma\tau}$ has
been shifted to decouple $\pi$ and $H$ from $\tilde b$. As a
result of this shift, the action for $\tilde b$ corresponds to a
scalar of mass $M^2l^2=45$, and is identified with the $k=1$ mode
in the second branch of mass eigenvalues.

\subsection{Antisymmetric Tensor Spectrum}

The action for a free massless complex ten-dimensional
antisymmetric tensor can be written as

\be S=\int d^{10}x\sqrt{-g}\left( -\frac{1}{2}
\del_m\bar{A}_{nk}(\del^mA^{nk}-\del^nA^{mk}-\del^kA^{nm})\right)\ee
Writing this in terms of five-dimensional components and making
use of the gauge conditions $\del^\alpha
A_{\alpha\beta}=\del^\alpha A_{\alpha b}=0$ gives

\bq S&=&\int d^{10}x\sqrt{-g}\left( -\frac{1}{2}
\left(\del_\mu\bar{A}_{\alpha\beta}\del^\mu
A^{\alpha\beta}+\del_\gamma\bar{A}_{\alpha\beta}\del^\gamma
A^{\alpha\beta}+ 6\bar{A}_{\alpha\beta}A^{\alpha\beta}\right)
\right.\n &-&\left(\del_\mu\bar{A}_{\nu\alpha}(\del^\mu
A^{\nu\alpha}-\del^\nu A^{\mu\alpha}) +
\del_\beta\bar{A}_{\mu\alpha}\del^\beta A^{\mu\alpha}+4\bar{A}_{\mu\alpha}A^{\mu\alpha}\right)\nonumber\\
&-&\left.\frac{1}{2} \left(\del_\mu\bar{A}_{\nu\rho}(\del^\mu
A^{\nu\rho}-\del^\nu A^{\mu\rho}-\del^\rho A^{\nu\mu})+
\del_\alpha\bar{A}_{\mu\nu}\del^\alpha A^{\mu\nu}\right)\right).
\label{ac}\eq The three lines of (\ref{ac}) correspond to scalar,
vector, and antisymmetric tensor fields on $AdS_5$, as is clear
when we perform the expansion in spherical harmonics:

\bq A_{\mu\nu}&=&\sum a_{\mu\nu}(x)Y(y),\qquad A_{\mu\alpha}=\sum
a_\mu(x)Y_\alpha(y),\n A_{\alpha\beta}&=&\sum
a(x)Y_{\alpha\beta}(y),\qquad B=\sum B(x)Y(y). \eq To reproduce
the equations of motion in \cite{Kim} we must add to (\ref{ac}) a
topological mass term

\be
S_{mass}=i\epsilon^{\alpha\beta\gamma\delta\epsilon}\bar{A}_{\alpha\beta}\p_\gamma
A_{\delta\epsilon}+
i\epsilon^{\mu\nu\rho\sigma\tau}\bar{A}_{\mu\nu}\p_\rho
A_{\sigma\tau}.\ee We can rewrite the antisymmetric tensor part of
(\ref{ac}) as a first-order system by introducing auxiliary fields
$B_{ab}$ and $\bar{B}_{ab}$: \bq S_{A}&=&\int
d^{10}x\sqrt{-g}\left(
-\frac{i}{2}\epsilon^{\mu\nu\rho\sigma\tau}\bar{B}_{\mu\nu}\p_\rho
A_{\sigma\tau}
+\frac{i}{2}\epsilon^{\mu\nu\rho\sigma\tau}B_{\mu\nu}\p_\rho\bar{A}_{\sigma\tau}
\right.\nonumber\\
&-&\left. 2\bar{B}_{\mu\nu}B^{\mu\nu}
-\frac{1}{2}\del_\alpha\bar{A}_{\mu\nu}\del^\alpha A^{\mu\nu} +
i\epsilon^{\mu\nu\rho\sigma\tau}\bar{A}_{\mu\nu}\p_\rho
A_{\sigma\tau}\right) \label{a7} \eq Changing the variables to

\be C=\hf\Box_y^{1/2}A+\Box_y^{-1/2}(B-A),\qquad \bar
C=\hf\Box_y^{1/2}\bar A-\Box_y^{-1/2}(\bar B-\bar A), \ee gives
standard mass terms with eigenvalues as in \cite{Kim}. From the
point of view of calculating the anomaly, it is clear that a
complex antisymmetric tensor in the first-order formalism is
equivalent to a real antisymmetric tensor in the second-order
formalism, as considered in section 5.

\section{Fermion Spectrum}

The action for the ten-dimensional spinor field is \be S=\int
d^{10}x\sqrt{-g}\left( {\bar{\hat\lambda}}\Gamma^mD_m\hat\lambda -
\frac{i}{2\cdot 5!}{\bar{\hat\lambda}}
\Gamma^{mnpqr}F_{mnpqr}\hat\lambda \right),\label{fa1} \ee We
choose the following representation of the $\Gamma$-matrices \bq
&&\Gamma^a=\sigma^1\otimes I_4\otimes \gamma^a,\quad
\Gamma^\alpha=-\sigma^2\otimes\tau^\alpha\otimes I_4 \nonumber\\
&&\{\Gamma_{{M}} ,\Gamma_{{N}}\} =2\eta_{{M}{N}}, \quad
\{\gamma_{{a}} ,\gamma_{{b}}\} =2\eta_{{a}{b}},\quad
\{\tau_{{\alpha}} ,\tau_{{\beta}}\} =2\delta_{{\alpha}{\beta}}
\nonumber \eq In this representation the matrix $\Gamma_{11}$ is
equal to \be \Gamma_{11} =\Gamma^{{0}}\cdots \Gamma^{{9}} = \left(
\begin{array}{cc}
I_{16}& 0\\ 0&  -I_{16}
\end{array}\right)
\ee The spinor is right handed, and the gravitino left-handed:
\be {\hat\lambda} =\frac12 (1-\Gamma_{11}){\hat\lambda}=\left(\begin{array}{c} 0\\
\lambda
\end{array}\right){\hat\psi_\mu} =\frac12 (1+\Gamma_{11}){\hat\psi_\mu}
=\left(\begin{array}{c} 0\\
\psi_\mu
\end{array}\right).\ee

The action (\ref{fa1}) becomes \be S=\int d^{10}x\sqrt{-g}\,
\bar{\lambda}\left( \gamma^aD_a +i\tau^\alpha D_\alpha +1\right)\l
. \ee Expanding $\lambda$ in spherical harmonics \be \lambda
=\sum_{k\ge 0}\left( \l_k^+(x)\Xi_k^+(y)+
\l_k^-(x)\Xi_k^-(y)\right),\qquad\tau^\alpha D_\alpha\Xi_k^\pm
=\mp i(k+\frac52 )\Xi_k^\pm , \ee we obtain the five-dimensional
action \be S=\int d^{5}x\sqrt{-g_a}\, \sum_{k\ge 0}\left(
\bar{\lambda}_k^+\left( \gamma^aD_a +k+\frac72\right)\l_k^+
+\bar{\lambda}_k^-\left( \gamma^aD_a
-k-\frac32\right)\l_k^-\right) \ee There is no need to add any
boundary term to this action, as the boundary conditions that we
imposed in Section 2 ensure that the classical action does not
vanish on shell.

The ten-dimensional action for the gravitino is \be S=\int
d^{10}x\sqrt{-g}\left(
{\bar{\hat\psi}}_m\Gamma^{mnp}D_n{\hat\psi}_p + \frac{i}{4\cdot
5!}{\bar{\hat\psi}}_m\Gamma^{mnp}\Gamma^{mnpqr}F_{mnpqr}\Gamma_n{\hat\psi}_p
\right).   \ee Rewriting this in terms of five-dimensional fields
gives \bq S&=&\int d^{10}x\sqrt{-g}\,\left( \bar{\psi}_\mu\left(
\gamma^{\mu\nu\rho}D_\nu\psi_\rho -i\gamma^{\mu\nu}\tau^\alpha
D_\nu\psi_\alpha \right. + i\gamma^{\mu\nu}\tau^\alpha
D_\alpha\psi_\nu+
\gamma^{\mu}\tau^{\alpha\beta} D_\alpha\psi_\beta -\gamma^{\mu\nu}\psi_\nu\right)\nonumber\\
&+& \bar{\psi}_\alpha\left(
-i\tau^{\alpha\beta\gamma}D_\beta\psi_\gamma
-i\gamma^{\mu\nu}\tau^\alpha D_\mu\psi_\nu
+\gamma^{\mu}\tau^{\alpha\beta}D_\beta\psi_\mu -\left.
\gamma^{\mu}\tau^{\alpha\beta} D_\mu\psi_\beta
+\tau^{\alpha\beta}\psi_\beta\right) \right) \eq As in \cite{Kim}
we fix the local supersymmetries by transforming away all modes of
$\tau\cdot\psi$ except the one proportional to the Killing spinor,
on which the eigenvalue of $i\tau\cdot D_y$ is $5/2l$. We perform
the decomposition

\be \label{phi} \psi_\mu=\p_\mu+{D^T_\mu\over D\cdot
D^T}D^T\cdot\psi+{1\over d}\gamma_\mu\gamma\cdot\psi, \ee where
$D^T_\mu=(\delta^\nu_\mu-\gamma_\mu\gamma^\nu/5)D_\nu$ is
$\gamma$-transverse so that

\be \label{const} \gamma\cdot\phi=D\cdot\phi=0. \ee If we put
$\psi_1=\sqrt{D\cdot D^T}D^T\cdot\psi$ and
$\psi_2=\gamma\cdot\psi$ then the change of variables
$\psi_\mu\rightarrow (\phi_\mu,\psi_1,\psi_2)$ has a trivial
Jacobian. The expansion in spherical harmonics is given by

\bq
&&\psi^T_{\alpha }=\sum\,  \psi^{I_T}(x)\Xi_{(\alpha )}^{I_T}(y)+\psi^{I_L}(x)D^T_\alpha\Xi(y)\nonumber\\
&&\psi_\mu=\sum\,  \psi_\mu(x)\Xi^{I_L}(y)\nonumber\\
&&\tau\cdot D\Xi_{(\alpha )}^{I_T} =m^{I_T}\Xi_{(\alpha )}^{I_T}
=\mp i(k+\frac72 )\Xi_{(\alpha )}^{I_T} ,\quad k\ge 1\n
&&\tau\cdot D\Xi^{I_L} =m^{I_L}\Xi^{I_L} =\mp i(k+\frac52
)\Xi^{I_L} ,\quad k\ge 1 \eq The action for $\phi_\mu$ decouples

\be S_\phi=\int d^{10}x\sqrt{-g}\,\bar\phi^\mu(\gamma\cdot
D+(m^{I_L}-1)/l)\phi_\mu,\ee but to remove the constraints
(\ref{const}) we must introduce Lagrange multiplier fields that
are equivalent to introducing a pair of ghosts with masses
$Ml=\sqrt{(m^{I_L}-1)^2+4}$ \cite{rs}. Diagonalising the
$(\psi_1,\psi_2,\psi^{I_T},\psi^{I_L})$ system, we get two spinor
fields that cancel the ghosts, and two spinor fields of mass
$Ml=3+m^{I_L}$, $m^{I_T}-1$, in agreement with the spectrum of
\cite{Kim}. Finally, in the case of the Killing spinor $\eta$, the
shift \be \psi_\mu\to \psi_\mu +\frac53 i\gamma_\mu\eta \nonumber
\ee gives the action for the massless gravitino and the mass
$-11/2l$ for $\eta$.

\section{Conclusions}

We have shown that the AdS/CFT conjecture for IIB String theory/
$\cal N$=4 Super-Yang-Mills theory passes the stringent test of
requiring that the Weyl anomlies of the two theories match at
sub-leading order. This generalises the leading order test of
Henningson and Skenderis but avoids perturbation theory in the
metric by working with an exact solution to the Einstein equation
in the bulk. At sub-leading order all the multiplets of IIB
Supergravity contribute to the boundary theory Weyl anomaly an
amount given by a universal formula that involves the
four-dimensional heat-kernel for conformally covariant operators.
The regularised sum of these contributions involves only those
operators that appear in the boundary theory resulting in the
matching of the anomalies. Our approach generalises to other
AdS/CFT correspondences.

\pagebreak

\begin{table}[b]
\begin{center}
\caption{Mass spectrum. The supermultiplets (irreps of U(2,2/4))
are labelled
   by the integer $p$. Note that the doubleton ($p=1$) does not appear in
the
   spectrum. The $(a,b,c)$ representation of $SU(4)$ has dimension
   $(a+1)(b+1)(c+1)(a+b+2)(b+c+2)(a+b+c+3)/12$, and a subscript $c$
indicates
   that the representation is complex. (Spinors are four component Dirac
   spinors in $AdS_5$).}
\label{spec} \vskip .3cm
  \begin{tabular}{|cccc|}
\hline

     Field  & $SO(4)$ rep$^{\rm n}$ & $SU(4)$ rep$^{\rm n}$ &
$\Delta-2$      \\

  \hline

$\phi^{(1)}$ & $(0,0)$ & $(0,p,0)$ & $p-2$,\quad $p\ge2$ \\
$\psi^{(1)}$ & $(\hf,0)$ & $(0,p-1,1)_c$ & $p-3/2$,\quad $p\ge2$ \\
$A_{\mu\nu}^{(1)}$ & $(1,0)$ & $(0,p-1,0)_c$ & $p-1$,\quad
$p\ge2$ \\
\hline $\phi^{(2)}$ & $(0,0)$ & $(0,p-2,2)_c$ & $p-1$,\quad
$p\ge2$
\\
$\phi^{(3)}$ & $(0,0)$ & $(0,p-2,0)_c$ & $p$,\quad $p\ge2$
\\
$\psi^{(2)}$ & $(\hf,0)$ & $(0,p-2,1)_c$ & $p-1/2$,\quad $p\ge2$ \\
$A_\mu^{(1)}$ & $(\hf,\hf)$ & $(1,p-2,1)$ & $p-1$,\quad $p\ge2$
\\
$\psi_\mu^{(1)}$ & $(1,\hf)$ & $(1,p-2,0)_c$ & $p-1/2$,\quad
$p\ge2$
\\
$h_{\mu\nu}$ & $(1,1)$ & $(0,p-2,0)$ & $p$,\quad $p\ge2$ \\

\hline
$\psi^{(3)}$ & $(\hf,0)$ & $(2,p-3,1)_c$ & $p-1/2$,\quad $p\ge3$ \\
$\psi^{(4)}$ & $(\hf,0)$ & $(0,p-3,1)_c$ & $p+1/2$,\quad $p\ge3$ \\
$A_\mu^{(2)}$ & $(\hf,\hf)$ & $(1,p-3,1)_c$ & $p$,\quad
$p\ge3$ \\
$A_{\mu\nu}^{(2)}$ & $(1,0)$ & $(2,p-3,0)_c$ & $p$,\quad $p\ge3$
\\
$A_{\mu\nu}^{(3)}$ & $(1,0)$ & $(0,p-3,0)_c$ & $p+1$,\quad
$p\ge3$ \\
$\psi_\mu^{(2)}$ & $(1,\hf)$ & $(1,p-3,0)_c$ & $p+1/2$,\quad
$p\ge3$
\\
\hline
$\phi^{(4)}$ & $(0,0)$ & $(2,p-4,2)$ & $p$,\quad $p\ge4$ \\

$\phi^{(5)}$ & $(0,0)$ & $(0,p-4,2)_c$ & $p+1$,\quad $p\ge4$
\\
$\phi^{(6)}$ & $(0,0)$ & $(0,p-4,0)$ & $p+2$,\quad $p\ge4$ \\
$\psi^{(5)}$ & $(\hf,0)$ & $(2,p-4,1)_c$ & $p+1/2$,\quad $p\ge4$ \\
$\psi^{(6)}$ & $(\hf,0)$ & $(0,p-4,1)_c$ & $p+3/2$,\quad $p\ge4$ \\
$A_\mu^{(3)}$ & $(\hf,\hf)$ & $(1,p-4,1)$ & $p+1$,\quad $p\ge4$
\\
\hline
  \end{tabular}
  \end{center}
\end{table}

\begin{table}[tbp]
\begin{center}
\caption{Anomaly coefficients of massive fields on $AdS_5$. Note
that the massive vector coefficient is $v_0+2s-2s_0$ where
$v_0,s,s_0$ are respectively, the coefficients for the 4d
gauge-fixed Maxwell operator, a conformally coupled scalar, and a
minimally coupled scalar. } \label{coeffs} \vskip .3cm
\begin{tabular}{|ccc|}
\hline
Field & $R_{{ij}}=0$: & Constant $R$:\\
        & $180 a_{2}/R_{ijkl}R^{ijkl}$ & $180a_2/R^2$ \\
\hline
$\phi$ & 1 & -1/12\\
$\psi$  & 7/2 & -11/12\\
$A_\mu$ & -11 & 29/3\\
$A_{\mu\nu}$&  33 & 19/4\\
$\psi_\mu$ & -219/2 & -61/4\\
$h_{\mu\nu}$ & 189 &747/4\\
\hline

   \end{tabular}
  \end{center}
\end{table}

\begin{table}[t]
\begin{center}
\caption{Decomposition of gauge fields for the massless
multiplet.} \label{ghosts} \vskip .3cm
\begin{tabular}{|c|cccc|}
\hline
Original field & Gauge fixed fields & $\Delta-2$ & $R_{{ij}}=0$:& Constant $R$:\\
    &  &  & $180 a_{2}/R_{ijkl}R^{ijkl}$ & $180a_2/R^2$\\
\hline
$A_\mu$ & $A_i$ & 1 & -11 & 29/3\\
({\bf 15} of $SU(4)$)       & $A_0$ & 2 & 1 & -1/12\\
         & $b_{FP}$, $c_{FP}$ & 2 & -1 & 1/12\\
\hline
$\psi_\mu$ & $\psi_i^{\rm irr}$ & 3/2 & -219/2 & -61/4\\
&            $\gamma^i\psi_i$ & 5/2 & 7/2 & -11/12\\
({\bf 4} of $SU(4)$)  & $\psi_0$ & 5/2 & 7/2 & -11/12\\
& $\lambda_{FP}$, $\rho_{FP}$ & 5/2 & -7/2 & 11/12\\
& $\sigma_{GF}$ & 5/2 & -7/2 & 11/12\\
\hline
$h_{\mu\nu}$ & $h_{ij}^{\rm irr}$ & 2 & 189 & 727/4\\
($SU(4)$ singlet) & $h_{0i}$ & 3 & -11 & 29/3\\
& $h_{00}$, $h_\mu^\mu$ & $\sqrt{12}$& 1 & -1/12\\
& $B^{FP}_0$,$C^{FP}_0$ &$\sqrt{12}$& -1 & 1/12\\
& $B^{FP}_i$,$C^{FP}_i$ & 3 & 11 & -29/3\\
\hline
   \end{tabular}
  \end{center}
\end{table}


\begin{thebibliography}{88}
\bibitem{Henningson} M. Henningson and K. Skenderis JHEP 9807 (1998),
023.

\bibitem{Maldacena} J. Maldacena, Adv.Theor.Math.Phys.2 (1998), 231.


\bibitem{bonora} L. Bonora, P. Cotta-ramusino and C. Reina,
Phys. Lett.  126B (1983) 305.




\bibitem{Duff1} M.J. Duff, Class. Quant. Grav. 11 (1994) 1387, and refs. therein.

\bibitem{testn} P. Mansfield and D. Nolland Phys.Lett.B495:435-439,
(2000).

\bibitem{us}
P. Mansfield and D. Nolland, JHEP 9907 (1999), 028.

\bibitem{us3}
P. Mansfield and D. Nolland, Phys.Lett.B515:192-196,2001.

\bibitem{d} D. Nolland, hep-th/0310169.

\bibitem{finiteness} D. Nolland, hep-th/0310201.

\bibitem{rs} D. Nolland, Phys.Lett.B485:308-310,2000.

\bibitem{witten}I.R.Klebanov and E.Witten, Nucl.Phys.B556 (1999),
89.

\bibitem{frolov}G.E.Arutyunov and S.A.Frolov, JHEP 9908 (1999),
024.

\bibitem{Kim}
H.J.~Kim, L.J.~Romans and P.~van Nieuwenhuizen, Phys. Rev. D32
(1985), 389.

\bibitem{Duff2} S.M. Christensen and M.J. Duff, Nucl.Phys.B 154
(1979)301.

\bibitem{Barv} A.O. Barvinsky and G.A. Vilkovisky, Phys. Rep.
119(1985)1.

\bibitem{mnu-prep} P. Mansfield, D. Nolland and T. Ueno, Phys.Lett. B565 (2003)
207-210.



\bibitem{strong} P. Mansfield, Nucl.Phys.B418:113-130,1994.


\end{thebibliography}
\end{document}